\documentclass[%
 reprint,
superscriptaddress,
 amsmath,amssymb,
prl,
]{revtex4-2}

\DeclareMathOperator{\sinc}{sinc}
\usepackage{graphicx}
\usepackage{dcolumn}
\usepackage{bm}

\begin{document}

\preprint{APS/123-QED}

\title{Quantifying Hidden Nonlinear Noise in Integrated Photonics}

\author{Ben M. Burridge}
\email{ben.burridge@bristol.ac.uk}
\affiliation{Quantum Engineering Technology Laboratories, University of Bristol, Bristol, United Kingdom}
\affiliation{Quantum Engineering Centre for Doctoral Training, Centre for Nanoscience \& Quantum Information, University of Bristol, Bristol, United Kingdom}
\author{Imad I. Faruque}
\affiliation{Quantum Engineering Technology Laboratories, University of Bristol, Bristol, United Kingdom}
\author{John G. Rarity}
\affiliation{Quantum Engineering Technology Laboratories, University of Bristol, Bristol, United Kingdom}
\author{Jorge Barreto}
\affiliation{Quantum Engineering Technology Laboratories, University of Bristol, Bristol, United Kingdom}

\date{\today}

\begin{abstract}
We present experimental and simulated results to quantify the impact of nonlinear noise in integrated photonic devices relying on spontaneous four-wave mixing. Our results highlight the need for design rule adaptations to mitigate the otherwise intrinsic reduction in quantum state purity. The best strategy in devices with multiple parallel photon sources is to strictly limit photon generation outside of the sources. Otherwise, our results suggest that purity can decrease below 40\%.
\end{abstract}

\maketitle


Nonlinear light-matter interactions are an essential tool for the generation of photon states, the foundation of quantum photonics research in areas of communication \cite{qcomms}, computation \cite{qcomp}, and metrology \cite{metrologyreview}. Furthermore, they enable key active technologies such as electro-optic modulators \cite{eoswitch} and all-optical switches \cite{opticswitch}. Here, we target the practical implementation of high purity heralded single-photon sources (HSPSs) \cite{theorymzirings, theorydualpulse, mzirings, dualpulse, intermodalpurity}. In isolation, they are expected to perform very well, but we question the impact of surrounding infrastructure in current generation circuits.

Making the assumption that the designated ``photon-source'' is where all, or at least most of the nonlinear interactions will occur, trivializes the effect of such occurrences in surrounding components. Previous works have investigated the consequences of spurious photon-generation and how they affect the outcome of their experiment \cite{Silverstonespiralnoise}, even speculating on how future circuits may avoid the penalties of such events.

Here, we build on this work to produce a method to determine the impact of nonlinear noise on targeted quantum states. Using the silicon-on-insulator (SOI) platform as an example, probabilistic photon generation occurs through the spontaneous four-wave mixing (SFWM) process, thanks to the $\chi^{(3)}$ of silicon. However, due to the size of single-mode waveguides on the SOI platform (500 x 220~$nm$); which prevents light coupling to higher order modes, SFWM is phase-matched over the telecom C-band and can occur in all the waveguides that we use. For the first time, we quantify the effect of nonlinear noise stemming from conventional photonic architectures and use our modeling to predict the implications on a circuit's ability to create targeted quantum states. This work will therefore underline the importance of minimizing the brightness of photon generation outside of photon sources and offer accurate predictions as to the performance we can expect from real-world quantum photonic circuits. Further, we validate our model using a fabricated photonic integrated circuit (PIC) which allows us to offer practical insights into future architectures of PICs.

Commonly \cite{samplingstates, teleport, imadrings, silverstonerings, highdimensional, arbtwoqubit, twoqubitentangle}, bright pump-fields are created using an off-chip laser, optically coupled to the integrated photonic circuit. Dense wavelength-division multiplexers (DWDMs) are typically used before the chip to isolate the desired pump spectrum, attenuating amplified spontaneous emission (ASE) at photon-counting wavelengths. Further, DWDMs can be used before detection to isolate specific wavelengths of single-photons and remove any residual pump-field with high (100~dB \cite{filteringintegrated}) extinction. Filtering the pump to this extent ensures that light reaching single-photon detectors is overwhelmingly due to nonlinear photon generation inside the circuit. Once the pump field is on-chip, so long as the nonlinear process of choice is phase-matched, photons will be created probabilistically.

Many integrated photonic experiments that utilize multiple photon sources \cite{integratedquantumphotonic, samplingstates, teleport, imadrings, silverstonerings, highdimensional, arbtwoqubit, twoqubitentangle} will distribute the pump to each source using fixed couplers or tunable interferometric devices such as Mach-Zehnder Interferometers (MZIs). MZIs are commonly four-port devices but can be cascaded to produce N-port devices \cite{largescaleswitches, photonicprocessor}, and are extremely useful for balancing optical power across each output-port. 
This design philosophy comes from the inherently scalable infrastructure of PICs. Quantum PICs, however, should generally not be treated the same. While devices can be designed independently of the surrounding infrastructure, we must still treat the quantum PIC as a complete system. Further, stacking these devices together means that the effects seen in \cite{Silverstonespiralnoise} will occur inside every MZI where this bright pump field is present, leading to unexpected behavior of the wider circuit. In practice, if spurious light mixes with our pure quantum state, this fundamentally limits the purity of all implementations of photon sources \cite{nonlineargeneration, singlephotonsources}.
\begin{figure*}[htbp]
\centering\includegraphics[width = \linewidth]{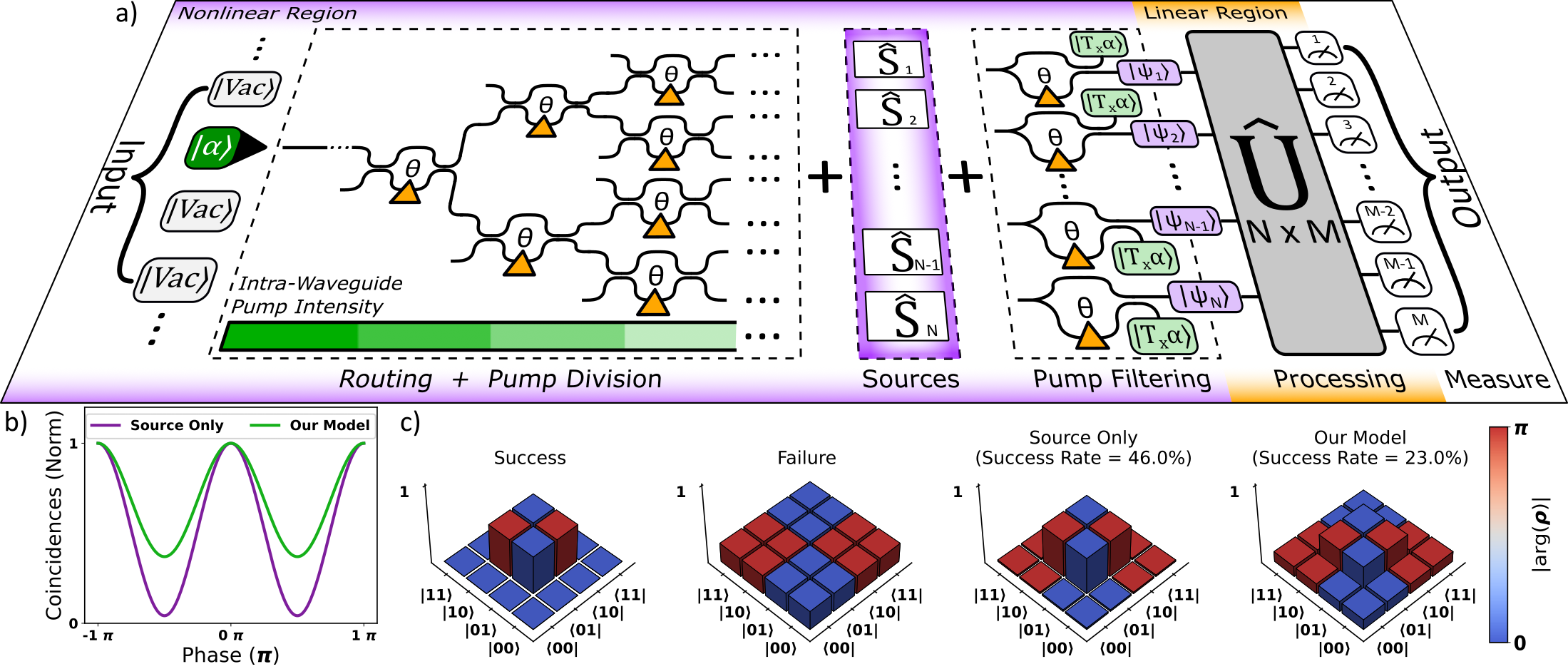}
\caption{a) Schematic representation of a generic quantum photonic circuit. Here we have the input of $|\alpha\rangle$, which propagates through a combination of components (governed by their total transmission $T_{x}$) before being removed. It is clear that there's a large region of the circuit where the $\chi^{(3)}$ interaction can take place outside of the sources, in contrast to the common assumption that photons are mostly generated by the sources. b) Simulated HOM interference fringes using our model, compared to the assumption that photons only come from the source. c) Simulation results for creating the $|\Psi^{-}\rangle$ state using two HSPSs. From left to right we include the density matrices ($\rho$) for the pure and mixed states resulting from successful and unsuccessful HOM interference, as well as the states that we can expect if photons only come from the source, or if we use our model. Here we use a simple ring resonator \cite{imadrings} as our source of comparison.}
\label{fig:Model}
\end{figure*}

There are many examples \cite{samplingstates, teleport, imadrings, highdimensional, arbtwoqubit, twoqubitentangle} where the source is considered as the sole nonlinear object when simulating an integrated quantum photonic experiment. This is a flawed assumption, as illustrated in Figure~\ref{fig:Model} by the distinction between linear and nonlinear regions. We present a more accurate model that exemplifies the importance of accounting for and minimizing spurious generation events as we push towards higher fidelity sources of quantum light \cite{nonlineargeneration, singlephotonsources, intermodalpurity, theorydualpulse, theorymzirings, mzirings, dualpulse}. We hope this model can be used to future-proof integrated photonic architectures by providing insight into the fundamental noise floor of engineered quantum states.

We start by modeling the propagation of light through a photonic circuit. The Hamiltonian of a waveguide with a $\chi^{(3)}$ interaction is \cite{asyfields, Silverstonespiralnoise}

\begin{eqnarray}
&&\hat{H} = \hat{H}_{L} + \hat{H}_{NL} \equiv \int{dk\hbar\omega_{k}\hat{a}^{\dagger}_{k}\hat{a}_{k}} - \label{eq:hamiltonian}
\\ &&\gamma_{0}\int{dk_{1}} dk_{2}dk_{3}dk_{4}\Phi(k_{1}, k_{2}, k_{3}, k_{4}, z)\hat{a}^{\dagger}_{k1}\hat{a}^{\dagger}_{k2}\hat{a}^{}_{k3}\hat{a}^{}_{k4} + h.c , \nonumber{}
\end{eqnarray}

\noindent where $\gamma_{0}$ is the effective nonlinear coupling constant, $k$ and $\omega$ are the angular wavenumber and frequency of each field respectively, with $z$ being geometric length, and $\hbar$ the reduced Planck constant. $\Phi(k_{1}, k_{2}, k_{3}, k_{4}, z)$ is the biphoton wavefunction containing energy-consevation ($\hbar\omega_{4}=\hbar\omega_{1}+\hbar\omega_{2}-\hbar\omega_{3}$) and the phase-matching condition \cite{nonlinoptics}

\begin{equation}
\phi(\Delta k, z) = \int^{L}_{0}e^{i\Delta kz}dz = L\sinc\left(\frac{\Delta kL}{2}\right)e^\frac{i\Delta kL}{2},
\label{eq:Phase-Match}
\end{equation}

\noindent with $\Delta k$ being the phase mismatch ($k_{3}+k_{4}-k_{1}-k_{2})$ between the four fields, and L the length of the interaction region. Given the input of a coherent state $|\alpha\rangle$ and the Hamiltonian in Eq.~(\ref{eq:hamiltonian}), we can map out the behavior of any chosen combination of components. A simple example would be to pump N different photon-sources in parallel and transfer the final state, resulting from all prior nonlinear interactions, into an NxN unitary ($\hat{U}$) that mediates their linear interactions (Fig.~\ref{fig:Model}). In this instance we use a combination of routing waveguides, MZIs to balance power between sources, the sources themselves, and asymmetric MZI (AMZI) spectral filters to remove $|\alpha\rangle$. Removal of $|\alpha\rangle$ effectively halts further $\chi^{(3)}$ interactions and we are left with the photon-states that have already been probabilistically generated. Additionally, removing $|\alpha\rangle$ quickly once it is no longer required is of great importance for mitigating nonlinear loss mechanisms \cite{photonabsorption}.

Figure~\ref{fig:Model} presents our treatment of a generic quantum photonic circuit, starting from an input coherent state $|\alpha\rangle$, which is linearly evolved throughout the circuit until it is removed. As for the nonlinear behavior, we use the evolved pump field to calculate the generated photon-states inside each component, which must then also propagate through the circuit. We use this model to predict the visibility of quantum interference we could expect from a realistic implementation (Fig.~\ref{fig:Model}b), which leads to a corresponding decrease in the success probability of creating Bell states using heralded single-photons (Fig.~\ref{fig:Model}c). It is worth noting that we present a success rate rather than the usual fidelity metric $F = \langle\Psi^-|\,\rho\,|\Psi^-\rangle$. We do this because distinguishable photons can still bunch out of a HOM interferometer, effectively increasing fidelity. Moreover, these random bunching events become more common for lower fidelity quantum interference.\\
We can examine the behavior of each component individually with an input of $|\alpha\rangle$, here our components can be narrowed down to just two categories: single-rail (routing waveguides, sources), and multi-rail (MZIs). To compare components, we first combine the phase-matching condition in Eq.~\ref{eq:Phase-Match} and the assumption that terms dependent on $\Delta k$ are approximately unity over our region of interest such that
\begin{eqnarray}
\phi(\Delta k, z) \approx L. \label{eq:pmatch_approx}
\end{eqnarray}
Implying that we can modulate the strength of the nonlinear interaction just by changing the length parameter $L$. Secondly, we move $\Phi(k_{1}, k_{2}, k_{3}, k_{4}, z)$ into the frequency domain, and integrate our biphoton wavefunction over our pump frequencies ($\omega_{3}$, $\omega_{4}$)

\begin{eqnarray}
L\int\Phi(\omega_{1}, \omega_{2}, \omega_{3}, \omega_{4}) d\omega_{3}d\omega_{4} = L\Phi(\omega_{s}, \omega_{i}),
\label{eq:integrated_bwf}
\end{eqnarray}
where $\omega_{s}$ and $\omega_{i}$ are the angular frequencies of the signal and idler fields. Finally, we normalize the biphoton wavefunction such that $\int{|\Phi(\omega_{s},\omega_{i})|^{2}}d\omega_{s}d\omega_{i} = 1$. We can then use Eq.~\ref{eq:hamiltonian} to write down the SFWM interaction to first-order as \cite{photongentheory}

\begin{eqnarray}
&&|\Psi_{SFWM}\rangle = \exp{\left(-\frac{i\hat{H}_{NL}}{\hbar}\right)}|vac\rangle, \\
&&|\Psi_{SFWM}\rangle \approx |vac\rangle + \frac{\gamma_{0}\alpha^{2}}{\sqrt{2}}L\Phi(\omega_{s},\omega_{i})\hat{a_{s}}^{\dagger}\hat{a_{i}}^{\dagger}|vac\rangle, \nonumber{}
\label{eq:SFWM_Int}
\end{eqnarray}
where $\alpha$ is the field amplitude of the coherent state, and $\hat{a_s}^{\dagger}$ and $\hat{a_i}^{\dagger}$ denote the signal and idler creation operators respectively. Higher order terms become necessary when experiments require the parallel generation of multiple photon pairs, but for our purposes the first-order case serves as an illustrative example. Using Eq.~\ref{eq:SFWM_Int}, it is easy to see that the SFWM interaction in a single waveguide with the pump input in a coherent state, signal/idler in the vacuum state
\begin{eqnarray}
|\Psi_{in}\rangle = |\alpha\rangle|vac\rangle, \nonumber{} \label{eq:input_state}
\end{eqnarray}
is given by
\begin{eqnarray}
\left(1 + \frac{\gamma_{0}\alpha^{2}}{\sqrt{2}}L^\prime_{x}\Phi_{x}(\omega_{s},\omega_{i})\hat{a_{s}}^{\dagger}\hat{a_{i}}^{\dagger}\right) |T_{x}\alpha\rangle|vac\rangle. \label{eq:Single_Rail}
\end{eqnarray}
For brevity, we have modified the geometric length $L$ of this waveguide according to
\begin{eqnarray}
L_xT_{x}^2 = L^\prime_x, \label{Loss_mod}
\end{eqnarray}
Where $T_{x}^2$ is a transmission co-efficient that accounts for insertion / propagation losses, and $L^\prime_{x}$ is therefore the effective geometric length of waveguide $x$, and $\Phi_{x}(\omega_{s},\omega_{i})$ is the associated biphoton wavefunction, as in Eq.~\ref{eq:integrated_bwf}.
\begin{figure*}[htbp]
\centering\includegraphics[width = \linewidth]{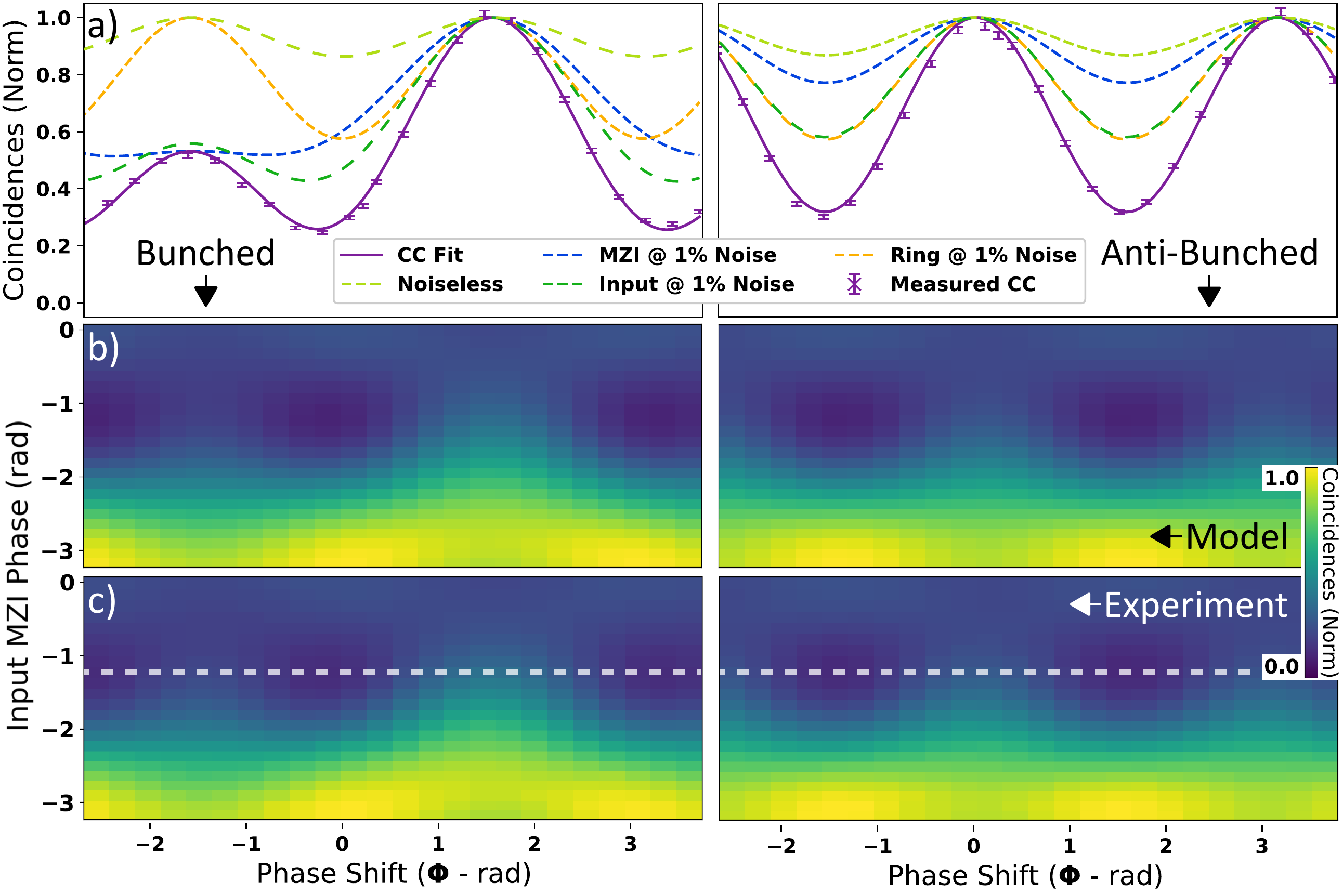}
\caption{We present simulations for biphoton interference fringes with varying types and proportions of noise present in an integrated photonic circuit. a) The effects of noise in individual components on the bunched (left) and anti-bunching photons (right). b) A fully modeled map of the interference expected from our quantum photonic circuit, for varying pump-splitting ratios between two distinct sources. c) A full experimental map of the same interference fringes from our fabricated circuit (Fig.~\ref{fig:Circuit}), where the dashed line indicates the data used in a).}
\label{fig:Noise_Effects}
\end{figure*}

For a 2x2 MZI, we start with input
\begin{eqnarray}
|\Psi_{in}\rangle = |\alpha\rangle|vac\rangle|vac\rangle. \nonumber{}
\end{eqnarray} 
Due to the multi-rail nature of the MZI, the first two states $|\alpha\rangle |vac\rangle$ are the input states of the pump for the upper and lower rail, respectively. The final $|vac\rangle$ state is the vacuum state of the signal/idler photons. This state is incident on a lossy coupler that splits the light across the two arms of the MZI,
\begin{eqnarray}
|\alpha\rangle|vac\rangle|vac\rangle \rightarrow |r\alpha\rangle|i\kappa\alpha\rangle|vac\rangle \label{simple_mzi}
\end{eqnarray} 
$r$ and $i\kappa$ are the modified reflection and transmission coefficients, such that 
\begin{eqnarray}
r^{2} + \kappa^{2} = T_{Split}^2. \nonumber{}
\end{eqnarray} 
Similar to Eq.~\ref{Loss_mod}, $T_{Split}^2$ accounts for the insertion loss of the couplers themselves, where a lossless coupler has $T_{Split}^2 = 1$.
Next, we need to consider linear propagation, and photon-generation (Eq.~\ref{eq:SFWM_Int}) in both arms leaving us with 
\begin{eqnarray}
&&\left(1 + \frac{(e^{i(k_pL_1 + \theta)}r\alpha)^2}{\sqrt{2}}L^\prime_1\Phi_{1}(\omega_{s},\omega_{i})\hat{a_{s}}^{\dagger}\hat{a_{i}}^{\dagger}\right) \otimes \nonumber{} \\ &&\left(1 - \frac{(e^{i(k_pL_2)}\kappa\alpha)^2}{\sqrt{2}}L^\prime_2\Phi_{2}(\omega_{s},\omega_{i})\hat{b_{s}}^{\dagger}\hat{b_{i}}^{\dagger}\right) \otimes \\ && |e^{i(k_pL_1 + \theta)}r\alpha\rangle|e^{i(k_pL_2)}i\kappa\alpha\rangle|vac\rangle. \nonumber{}
\end{eqnarray} 
Here we need to account for the phase-shift inside the MZI, to do this we include a tunable phase shift $\theta$, and propagation constant $k_pL_{1,2}$. As in Eq.~\ref{eq:Single_Rail}, $L^\prime_{1,2}$ are the effective geometric lengths of the individual waveguides between the couplers, and $k_p$ is the pump wavenumber. Additionally, $\hat{a}_{s,i}^{\dagger}$ and $\hat{b}_{s,i}^{\dagger}$ indicate photons generated in the upper and lower arms, respectively.\\
Finally, the state is incident on a second lossy coupler, that again splits the light according to Eq.~\ref{simple_mzi}, which we concatenate to get
\begin{subequations}
\begin{eqnarray}
\left(1 + \frac{\gamma_{0}\alpha^{2}}{\sqrt{2}}\Phi(\omega_{s},\omega_{i})\hat{S}_{FWM}\right)\label{eq:MZI_Linear}|T_{1}\alpha\rangle |T_{2}\alpha\rangle|vac\rangle,
\end{eqnarray} 
where we make the general assumption that $\Phi_{1}(\omega_{s},\omega_{i}) \equiv \Phi_{2}(\omega_{s},\omega_{i})$ with the final result of the nonlinear interaction ($\hat{S}_{FWM}$) given by

\begin{eqnarray}
&&(L^\prime_{1}e^{2i(k_pL_{1} + \theta)}r^{4} + L^\prime_{2}e^{2i(k_pL_{2})}\kappa^{4})\hat{a_{s}}^{\dagger}\hat{a_{i}}^{\dagger} - \label{eq:MZI_Nonlinear} \\ &&r^{2}\kappa^{2}(L^\prime_{1}e^{2i(k_pL_{1} + \theta)} +  L^\prime_{2}e^{2i(k_pL_{2})})\hat{b_{s}}^{\dagger}\hat{b_{i}}^{\dagger} + \nonumber{} \\ &&i\kappa r(L^\prime_{1}e^{2i(k_pL_{1} + \theta)}r^{2} - L^\prime_{2}e^{2i(k_pL_{2})}\kappa^{2})(\hat{a_{s}}^{\dagger}\hat{b_{i}}^{\dagger} + \hat{b_{s}}^{\dagger}\hat{a_{i}}^{\dagger}),\nonumber{}
\end{eqnarray}

\noindent and the transmission of the upper ($T_{1}$) and lower ($T_{2}$) arms

\begin{eqnarray}
&&T_{1} = \left(e^{i(k_pL_{1} + \theta)}r^{2} - e^{i(k_pL_{2})}\kappa^{2}\right), \label{eq:MZI_Linear_Trans} \\
&&T_{2} = i\kappa r\left(e^{i(k_pL_{1} + \theta)} + e^{i(k_pL_{2})}\right). \nonumber{}
\end{eqnarray}
\end{subequations}
It is intuitive to see how this scales up, as for every rail you add to the circuit, you gain an extra mode in both the pump, and signal/idler fields.

Using the results of Eq.~(\ref{eq:hamiltonian}-\ref{eq:MZI_Linear_Trans}), we can create a circuit model to accurately predict the photon-states that are created around a photon-source up until $|\alpha\rangle$ is removed. Once $|\alpha\rangle$ is removed the remainder of the model can be simplified to a linear scattering problem with negligble nonlinear effects. By tuning the length of each component, we can effectively control its contribution to the final state. This lets us see the exact contribution of the components that are not designated photon-sources, which we present in Figure~\ref{fig:Noise_Effects}.

\begin{figure*}[htbp]
\centering\includegraphics[width = \linewidth]{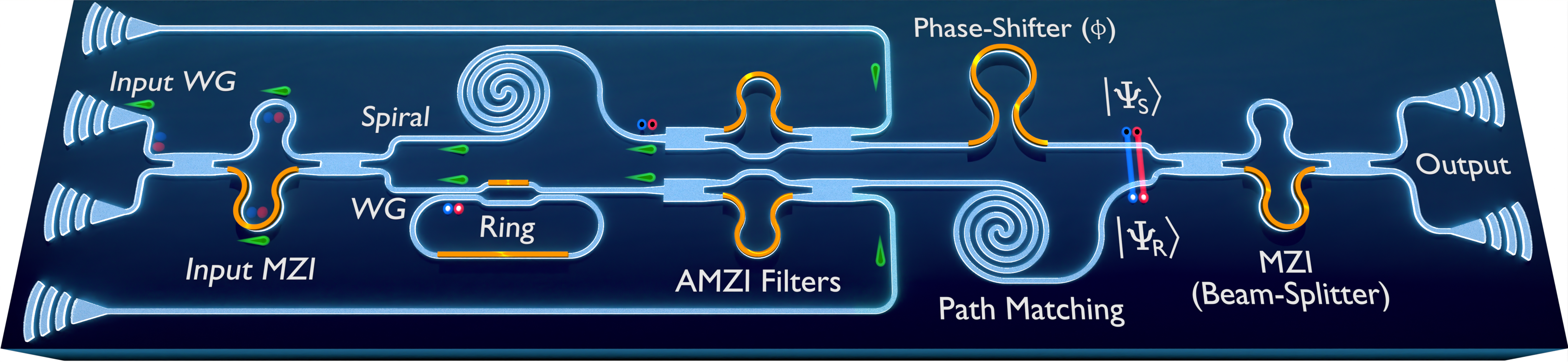}
\caption{We present the photonic circuit used to verify the model and its predictions. The pump (green) is input through a grating coupler on the left. It generates photon-pairs (represented by the red and the blue spheres) everywhere that it is present. The centres of the red/blue spheres represent the origin of the photon-pair; clear – spurious pairs, black – Spiral, white – Ring. The pump is removed, and the pairs are brought together on a final beam-splitter, and collected out of both grating coupler outputs. The phase-shifter ($\phi$) is used to control the relative phase of the $|\Psi_{S}\rangle$ and $|\Psi_{R}\rangle$, and the input MZI is used to change the relative power between the two sources.
}
\label{fig:Circuit}
\end{figure*}

We can also verify these observations experimentally using an integrated photonic circuit, our circuit contains two sources within an interferometer as described elsewhere \cite{SpET} (Fig.~\ref{fig:Circuit}). We measure two-fold coincidences between the signal and idler on a single output arm (bunched) and across both output arms (anti-bunched), signal and idlers are separated off-chip using standard DWDMs. We scan the phase-shifter ($\phi$) in Fig.~\ref{fig:Circuit} to obtain a biphoton interference fringe, resulting in the experimental data in Fig.~\ref{fig:Noise_Effects}. Our pump is a pulsed laser operating at 1550~nm, with a bandwidth of 260~pm and a repetition rate of 50~MHz, our resonant source is operated at a bandwidth of 60~pm.
Figure~\ref{fig:Noise_Effects}a demonstrates the effect of introducing noise inside specific components. We have separated the detection events where photons preferentially bunch and anti-bunch, making it easy to see the interference discrepancies in both cases, and an additional asymmetry present in the bunching events. Experimentally this is trivial when we measure both output ports, because coincidences are naturally separated into these two categories. Additionally, we present experimental measurements of our fabricated photonic circuit, and plot this alongside theoretical fits. The model represents the data well, and it becomes easy to see how the sources of noise could be combined to create the measured interference pattern. Taking the analysis further, we can completely map the behavior of our quantum circuit using this approach, presented experimentally (Fig.~\ref{fig:Noise_Effects}c) alongside the model using estimated component lengths that we adjust using typical device insertion losses (Fig.~\ref{fig:Noise_Effects}b). Performing residual analysis on this data, we see mean residuals of $1.1 \pm 0.8~\%$ and $0.6 \pm 0.9~\%$ for the anti-bunching and bunching events, respectively.\\
When considering the probabilistic generation of photon-pairs, the mitigation of multi-pair terms is crucial if aiming for high interference fidelity and the detection mechanism is non-number resolving. We made sure that, at least when considering the spurious noise of our circuit, the interference patterns we have measured are solely due to the design of the circuit. Decreasing the incident on-chip pump power by more than an order of magnitude (See Supplemental Material) leads to no overall change in interference visibility.

\begin{figure*}[htbp]
\centering\includegraphics[width = \linewidth]{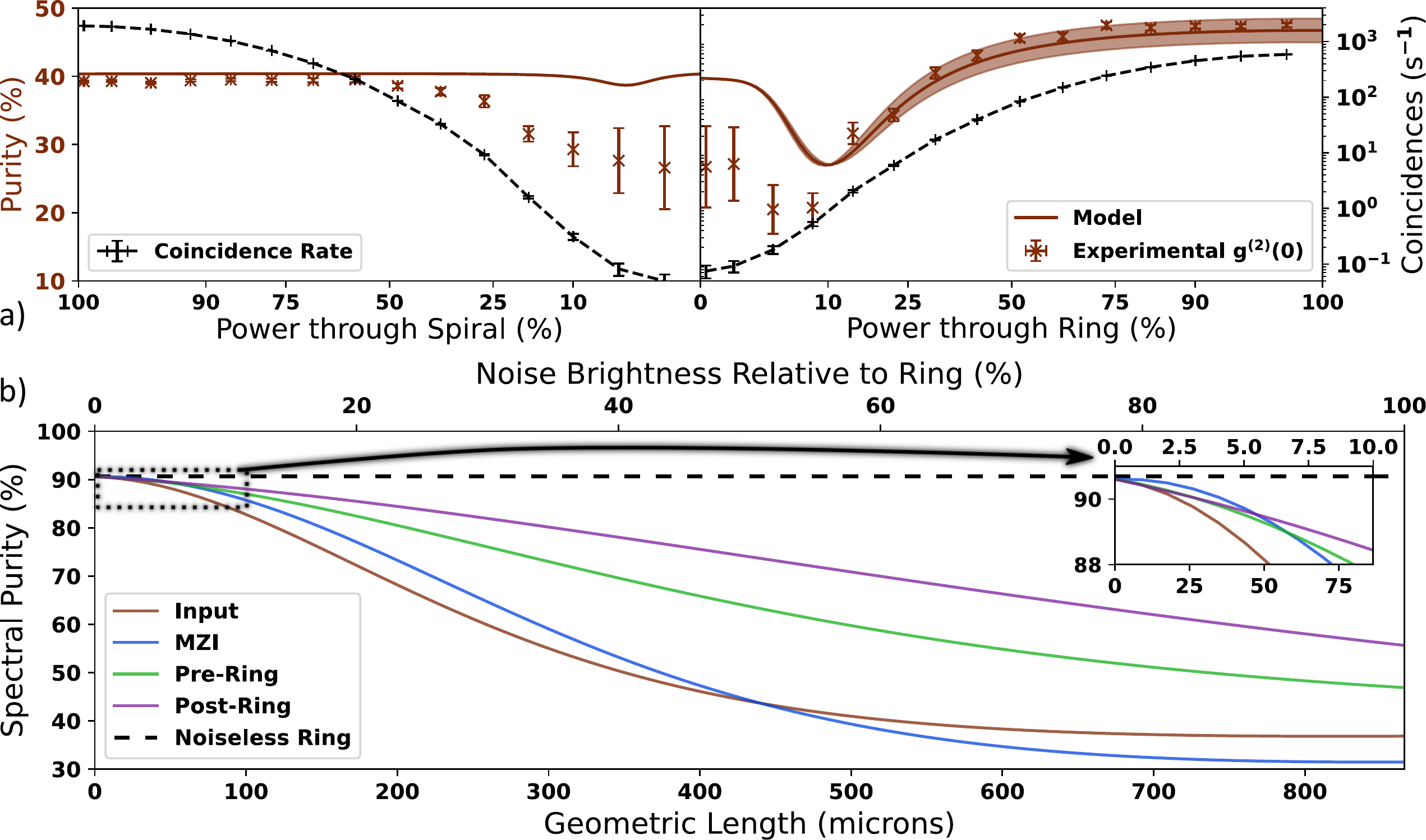}
\caption{a) $g^{(2)}(0)$ measurements for the spiral ($|\Psi_{S}\rangle$ - left) and the ring ($|\Psi_{R}\rangle$ - right) with a filter width of 1.3~nm, here the spiral acts as our reference source. The shaded regions indicate the confidence in our model due to source brightness estimations, we also track our coincidence rates to make inferences about noise present in the measurements. b) Purity decay of our pure photon source with the lengths of different components as in Fig.~\ref{fig:Circuit} for a filter width of 1.3~nm. 100~\% relative brightness means that spurious generation is equally likely to intentional generation. Pre-ring and Post-ring denotes whether the noise is generated before or after the ring, and the trends assume a 50:50 splitting ratio. The insert is a zoomed in portion that highlights the earlier purity trends. See Supplemental Material for further details.}
\label{fig:Purity_Decay}
\end{figure*}

This modeling of our circuit (Fig.~\ref{fig:Circuit}) also allows us to gain insight into the purity of the actual quantum state that we generate. This leads us to some interesting properties of the state (Fig.~\ref{fig:Purity_Decay}a), where we see the purity of the quantum state out of the ring is far below what we might have initially expected. Here our only free-parameter is the MZI splitting ratio, and in the typical operating scenario (50:50 for power division between multiple sources) the purity is lower than the one expected from a single source. This trend highlights the need to treat MZI's with care when using them to balance sources if the MZI itself is a source of nonlinearity. We use unheralded $g^{(2)}(0)$ measurements to experimentally characterize the spectral purity of a stream of photons. In our unheralded $g^{2}(0)$ measurements we only filter and detect the signal photons, but in standard operation idler photons would receive the same treatment to remove as much noise as possible. The byproduct of this is a slight purity increase beyond what we present here.
Figure~\ref{fig:Purity_Decay}a presents the purity of $|\Psi_{R}\rangle$ which we are most interested in, and also that of $|\Psi_{S}\rangle$. The purity of $|\Psi_{S}\rangle$ should be approximately unchanged with the MZI splitting ratio, although we do estimate some influence from the anti-bunched terms still propagating through the ring. From this comparison, we observe that as the coincidence rate drops we also begin to see deviation from the model in the spiral and ring case. Thankfully, by comparing the two sources, we notice that they decay to the same purity and can therefore infer that the measured purity at lower coincidence rates is limited by noise.
Figure~\ref{fig:Purity_Decay}b investigates the effects of surrounding component effective lengths on the quantum state purity we can expect from $|\Psi_{R}\rangle$. We present the strength of our ring relative to an equivalent length of waveguide for the filtering conditions used, in this instance (Fig.~\ref{fig:Purity_Decay}) we used a standard DWDM filter of 1.3~nm in bandwidth. In the presence of spurious photon generation, even conventionally short devices O(100~$\mu m$) can present disastrous consequences for the purity of the final state, with initial routing waveguides and splitting MZIs having the biggest impact.

Here, we have modeled a photonic circuit of two distinct sources, but this is a best case scenario. As the quantum photonics community aims to scale the number of sources, we also scale the size of our pump-distribution trees. Consequently, the earlier pump-splitters will become a larger source of spurious noise as they will contribute increasingly to the final quantum state. This has larger implications for the creation of one of our most fundamental resources for quantum communication \cite{qinternet} and quantum information processing \cite{qcompfusion}, the efficient creation of Bell pairs. 

\begin{figure*}[htbp]
\centering\includegraphics[width = \linewidth]{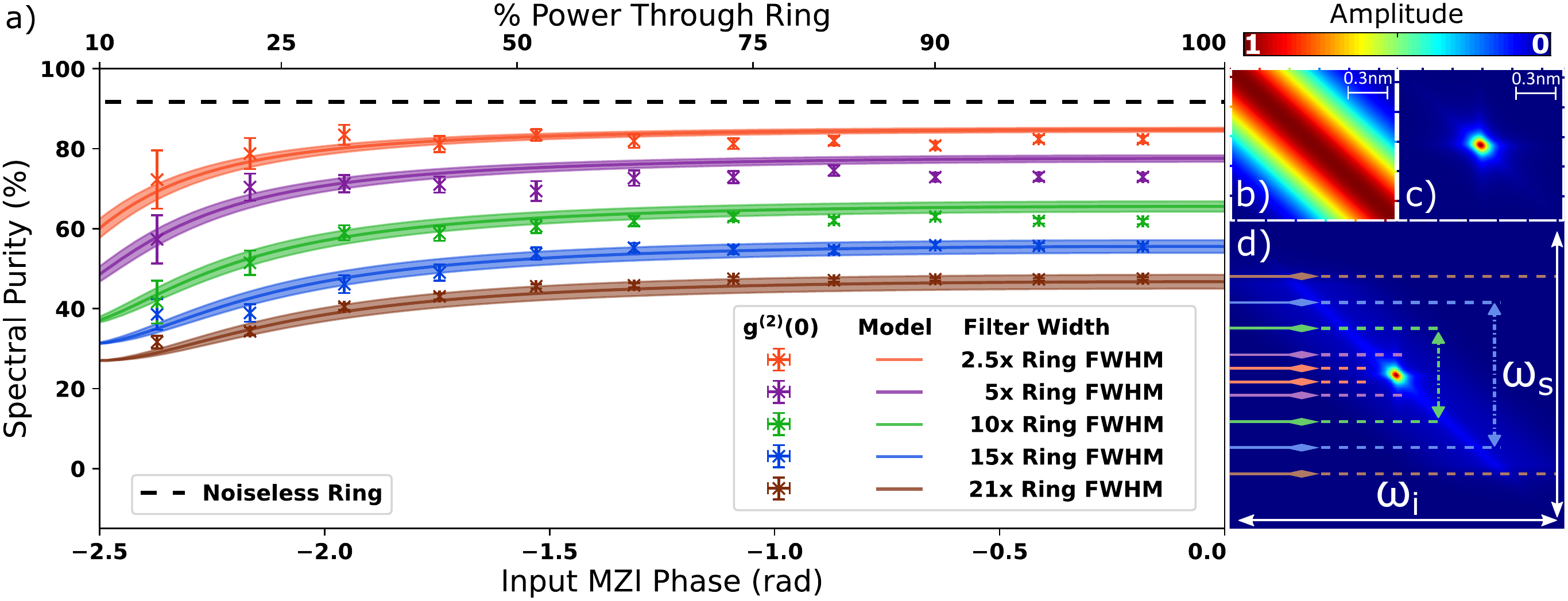}
\caption{a) Effect of the size of a spectral filter relative to the photon source bandwidth (60~$pm$) on spectral purity. Shaded regions indicate the confidence in the model due to source brightness estimations. Simulated biphoton wavefunctions of a b) simple waveguide and c) our resonator. d) Filtering conditions on the simulated compound biphoton wavefunction (colors correspond to the legend in a) ) at $|\Psi_{R}\rangle$ (Fig.~\ref{fig:Circuit}).}
\label{fig:Purity_Filtering}
\end{figure*}

Beyond employing MZI's that use some combination of multimode waveguides to suppress photon generation, and long adiabatic tapers to avoid exciting higher order modes, a simple solution is to employ filtering to specifically remove the noise from targeted wavelengths. We explore two modes of operation; filtering tightly around the photons before detection (behavior explored in Fig.~\ref{fig:Purity_Filtering} and \cite{imadg2}), preemptively filtering any spurious photons that were created in the components preceding your designated photon source. 
As we can see from our model in Figure~\ref{fig:Purity_Filtering}, filtering tightly before detection is not a perfect solution, and using the input MZI in a 50:50 configuration always leads to purity degradation of between 1-3\%. If the sources in question require anything other than this splitting ratio, their spectral purity will diverge, highlighting a need for tolerant source designs. This is concerning as here we are presenting the best case scenario through what is effectively a two source experiment, as we attempt to move towards large numbers of high purity sources this problem can only get worse. Although it is possible to recover most of the quantum state purity when filtering, there is a limit to how tightly it can be filtered before affecting the heralding efficiency of your source. If we filter immediately before the photon source instead, we can filter out all the noise across the wavelengths of interest before going on to generate our pure quantum states of light, creating a "wall-plug" source \cite{filterpresource}. Consequently, this would decouple the purity dependence on the power-splitting ratio of the MZI, and in general on the amount waveguide surrounding the source.

Experimentally, we characterize the effect of filtering after the source using unheralded $g^{(2)}(0)$ measurements with progressively tighter filtering bandwidths (Fig.~\ref{fig:Purity_Filtering}). The counterpart to our measurements could form a useful comparison between the two techniques to demonstrate the benefit of filtering on-chip prior to the photon source in a real-world circuit, \cite{mzirings} does this intrinsically but also filters very tightly for the $g^{2}(0)$ measurements. Alternatively, this information could also be captured using stimulated emission tomography \cite{SET}, but due to the broadness, and relatively low peak intensity of the non-resonant source, would require very high SNRs typical of time-resolved single photon detection \cite{photonspectrum}.
Our measurements show relatively good agreement with the model, with at worst a 2 - 3\% discrepancy for the more stringent filtering scenarios, which could be due to additional sources of noise \cite{realg2}. Here we have not characterized the phase profile or chirp of the pump \cite{imadg2}, as well as any contributions from raman scattering in fibre \cite{noisesources} or inelastic scattering in the Si waveguides \cite{inelasticscattering}, but from Figure~\ref{fig:Purity_Filtering} we can easily see that by far the largest contributor is spurious SFWM.
A further point of interest could be the accuracy of unheralded $g^{2}(0)$ measurements as a definitive method of purity characterization, when in standard operation we filter our data through heralding. This would effectively mitigate events where, on exiting the pump distribution section, spurious pairs have anti-bunched. Regardless, they serve perfectly well for model verification, but simulations without these anti-bunched spurious pairs are included in the supplementary.

To address our observations we propose a design rule change in the composition of integrated photonic circuits. It is clear that we risk bottlenecking the purity of our photon sources if we do not actively suppress the noise generated outside of them. Previous works \cite{Silverstonespiralnoise} modeled the effect of noise on the bunching and anti-bunching of photon pairs inside an MZI of fixed splitting ratio. Here, we extend that to a more general treatment, expanding beyond biphoton interference, and discussing the impact on the quantum state itself. We have shown that moving to resonant sources alone is not enough to suppress the impact of this noise. Wider, low-loss waveguide interconnects could act as a means of noise suppression by increasing the effective area of the mode, but with the added risk of exciting higher-order modes. Simply filtering allows us to retain the same infrastructure that we already have and it is instead the quality of the filter that determines the noise floor. Additionally, as we have mentioned SFWM can occur in any component so long as the process is phase-matched, this includes filters like the AMZI. These reliable filtering components are also a source of spurious noise, and typically operate in a transmit/reject regime for generated photons and pump, respectively, which means they produce streams of bunched photons. Therefore, any future implementations should consider the contributions from the source filter network to be the effective noise floor of the circuit, and as such their footprint must be minimized. This sort of noise floor consideration could be a limiting factor for sources that already implement AMZI filtering as part of their design \cite{theorymzirings, mzirings}, potentially giving pump-isolating designs like \cite{linearuncouple, linearuncouple_parasitic} an advantage. The need to minimize experimental noise floors underlines the demand for compact filters, which would allow us to efficiently isolate the source from any spurious FWM.

We envision this work to be fundamental in the engineering of low-noise integrated photonic circuits. We have presented a complete model of the quantum interference in our fabricated circuit using only waveguide lengths and estimated losses to accurately recreate the behavior, which in the future will drastically reduce the discrepancies between expectation and reality. Further, we have verified our model using interference patterns, and also the measured quantum state purity of our photons which provides additional confidence in our simulated quantum states. Finally, we offer solutions that allow the retention of quantum state purity and have contextualized this using the implementations of multiple different source designs. As we move towards larger and larger circuits with arrays of photon sources, the impact of spurious noise must be mitigated wherever possible, as the magnitude of such noise will only grow with scale. Our work addresses this limitation of integrated nonlinear circuits and offers solutions that will lead to predictable, high-performing, large-scale quantum PICs.


\bibliography{MAIN}

\begin{thebibliography}{40}%
\makeatletter
\providecommand \@ifxundefined [1]{%
 \@ifx{#1\undefined}
}%
\providecommand \@ifnum [1]{%
 \ifnum #1\expandafter \@firstoftwo
 \else \expandafter \@secondoftwo
 \fi
}%
\providecommand \@ifx [1]{%
 \ifx #1\expandafter \@firstoftwo
 \else \expandafter \@secondoftwo
 \fi
}%
\providecommand \natexlab [1]{#1}%
\providecommand \enquote  [1]{``#1''}%
\providecommand \bibnamefont  [1]{#1}%
\providecommand \bibfnamefont [1]{#1}%
\providecommand \citenamefont [1]{#1}%
\providecommand \href@noop [0]{\@secondoftwo}%
\providecommand \href [0]{\begingroup \@sanitize@url \@href}%
\providecommand \@href[1]{\@@startlink{#1}\@@href}%
\providecommand \@@href[1]{\endgroup#1\@@endlink}%
\providecommand \@sanitize@url [0]{\catcode `\\12\catcode `\$12\catcode
  `\&12\catcode `\#12\catcode `\^12\catcode `\_12\catcode `\%12\relax}%
\providecommand \@@startlink[1]{}%
\providecommand \@@endlink[0]{}%
\providecommand \url  [0]{\begingroup\@sanitize@url \@url }%
\providecommand \@url [1]{\endgroup\@href {#1}{\urlprefix }}%
\providecommand \urlprefix  [0]{URL }%
\providecommand \Eprint [0]{\href }%
\providecommand \doibase [0]{https://doi.org/}%
\providecommand \selectlanguage [0]{\@gobble}%
\providecommand \bibinfo  [0]{\@secondoftwo}%
\providecommand \bibfield  [0]{\@secondoftwo}%
\providecommand \translation [1]{[#1]}%
\providecommand \BibitemOpen [0]{}%
\providecommand \bibitemStop [0]{}%
\providecommand \bibitemNoStop [0]{.\EOS\space}%
\providecommand \EOS [0]{\spacefactor3000\relax}%
\providecommand \BibitemShut  [1]{\csname bibitem#1\endcsname}%
\let\auto@bib@innerbib\@empty
\bibitem [{\citenamefont {Joshi}\ \emph {et~al.}(2020)\citenamefont {Joshi},
  \citenamefont {Aktas}, \citenamefont {Wengerowsky}, \citenamefont
  {Lon{\v{c}}ari{\'c}}, \citenamefont {Neumann}, \citenamefont {Liu},
  \citenamefont {Scheidl}, \citenamefont {Lorenzo}, \citenamefont {Samec},
  \citenamefont {Kling} \emph {et~al.}}]{qcomms}%
  \BibitemOpen
  \bibfield  {author} {\bibinfo {author} {\bibfnamefont {S.~K.}\ \bibnamefont
  {Joshi}}, \bibinfo {author} {\bibfnamefont {D.}~\bibnamefont {Aktas}},
  \bibinfo {author} {\bibfnamefont {S.}~\bibnamefont {Wengerowsky}}, \bibinfo
  {author} {\bibfnamefont {M.}~\bibnamefont {Lon{\v{c}}ari{\'c}}}, \bibinfo
  {author} {\bibfnamefont {S.~P.}\ \bibnamefont {Neumann}}, \bibinfo {author}
  {\bibfnamefont {B.}~\bibnamefont {Liu}}, \bibinfo {author} {\bibfnamefont
  {T.}~\bibnamefont {Scheidl}}, \bibinfo {author} {\bibfnamefont {G.~C.}\
  \bibnamefont {Lorenzo}}, \bibinfo {author} {\bibfnamefont
  {{\v{Z}}.}~\bibnamefont {Samec}}, \bibinfo {author} {\bibfnamefont
  {L.}~\bibnamefont {Kling}}, \emph {et~al.},\ }\bibfield  {title} {\bibinfo
  {title} {A trusted node--free eight-user metropolitan quantum communication
  network},\ }\href@noop {} {\bibfield  {journal} {\bibinfo  {journal} {Science
  advances}\ }\textbf {\bibinfo {volume} {6}},\ \bibinfo {pages} {eaba0959}
  (\bibinfo {year} {2020})}\BibitemShut {NoStop}%
\bibitem [{\citenamefont {Ladd}\ \emph {et~al.}(2010)\citenamefont {Ladd},
  \citenamefont {Jelezko}, \citenamefont {Laflamme}, \citenamefont {Nakamura},
  \citenamefont {Monroe},\ and\ \citenamefont {O’Brien}}]{qcomp}%
  \BibitemOpen
  \bibfield  {author} {\bibinfo {author} {\bibfnamefont {T.~D.}\ \bibnamefont
  {Ladd}}, \bibinfo {author} {\bibfnamefont {F.}~\bibnamefont {Jelezko}},
  \bibinfo {author} {\bibfnamefont {R.}~\bibnamefont {Laflamme}}, \bibinfo
  {author} {\bibfnamefont {Y.}~\bibnamefont {Nakamura}}, \bibinfo {author}
  {\bibfnamefont {C.}~\bibnamefont {Monroe}},\ and\ \bibinfo {author}
  {\bibfnamefont {J.~L.}\ \bibnamefont {O’Brien}},\ }\bibfield  {title}
  {\bibinfo {title} {Quantum computers},\ }\href@noop {} {\bibfield  {journal}
  {\bibinfo  {journal} {nature}\ }\textbf {\bibinfo {volume} {464}},\ \bibinfo
  {pages} {45} (\bibinfo {year} {2010})}\BibitemShut {NoStop}%
\bibitem [{\citenamefont {Polino}\ \emph {et~al.}(2020)\citenamefont {Polino},
  \citenamefont {Valeri}, \citenamefont {Spagnolo},\ and\ \citenamefont
  {Sciarrino}}]{metrologyreview}%
  \BibitemOpen
  \bibfield  {author} {\bibinfo {author} {\bibfnamefont {E.}~\bibnamefont
  {Polino}}, \bibinfo {author} {\bibfnamefont {M.}~\bibnamefont {Valeri}},
  \bibinfo {author} {\bibfnamefont {N.}~\bibnamefont {Spagnolo}},\ and\
  \bibinfo {author} {\bibfnamefont {F.}~\bibnamefont {Sciarrino}},\ }\bibfield
  {title} {\bibinfo {title} {Photonic quantum metrology},\ }\href@noop {}
  {\bibfield  {journal} {\bibinfo  {journal} {AVS Quantum Science}\ }\textbf
  {\bibinfo {volume} {2}},\ \bibinfo {pages} {024703} (\bibinfo {year}
  {2020})}\BibitemShut {NoStop}%
\bibitem [{\citenamefont {Eltes}\ \emph {et~al.}(2020)\citenamefont {Eltes},
  \citenamefont {Villarreal-Garcia}, \citenamefont {Caimi}, \citenamefont
  {Siegwart}, \citenamefont {Gentile}, \citenamefont {Hart}, \citenamefont
  {Stark}, \citenamefont {Marshall}, \citenamefont {Thompson}, \citenamefont
  {Barreto} \emph {et~al.}}]{eoswitch}%
  \BibitemOpen
  \bibfield  {author} {\bibinfo {author} {\bibfnamefont {F.}~\bibnamefont
  {Eltes}}, \bibinfo {author} {\bibfnamefont {G.~E.}\ \bibnamefont
  {Villarreal-Garcia}}, \bibinfo {author} {\bibfnamefont {D.}~\bibnamefont
  {Caimi}}, \bibinfo {author} {\bibfnamefont {H.}~\bibnamefont {Siegwart}},
  \bibinfo {author} {\bibfnamefont {A.~A.}\ \bibnamefont {Gentile}}, \bibinfo
  {author} {\bibfnamefont {A.}~\bibnamefont {Hart}}, \bibinfo {author}
  {\bibfnamefont {P.}~\bibnamefont {Stark}}, \bibinfo {author} {\bibfnamefont
  {G.~D.}\ \bibnamefont {Marshall}}, \bibinfo {author} {\bibfnamefont {M.~G.}\
  \bibnamefont {Thompson}}, \bibinfo {author} {\bibfnamefont {J.}~\bibnamefont
  {Barreto}}, \emph {et~al.},\ }\bibfield  {title} {\bibinfo {title} {An
  integrated optical modulator operating at cryogenic temperatures},\
  }\href@noop {} {\bibfield  {journal} {\bibinfo  {journal} {Nature Materials}\
  }\textbf {\bibinfo {volume} {19}},\ \bibinfo {pages} {1164} (\bibinfo {year}
  {2020})}\BibitemShut {NoStop}%
\bibitem [{\citenamefont {Shcherbakov}\ \emph {et~al.}(2015)\citenamefont
  {Shcherbakov}, \citenamefont {Vabishchevich}, \citenamefont {Shorokhov},
  \citenamefont {Chong}, \citenamefont {Choi}, \citenamefont {Staude},
  \citenamefont {Miroshnichenko}, \citenamefont {Neshev}, \citenamefont
  {Fedyanin},\ and\ \citenamefont {Kivshar}}]{opticswitch}%
  \BibitemOpen
  \bibfield  {author} {\bibinfo {author} {\bibfnamefont {M.~R.}\ \bibnamefont
  {Shcherbakov}}, \bibinfo {author} {\bibfnamefont {P.~P.}\ \bibnamefont
  {Vabishchevich}}, \bibinfo {author} {\bibfnamefont {A.~S.}\ \bibnamefont
  {Shorokhov}}, \bibinfo {author} {\bibfnamefont {K.~E.}\ \bibnamefont
  {Chong}}, \bibinfo {author} {\bibfnamefont {D.-Y.}\ \bibnamefont {Choi}},
  \bibinfo {author} {\bibfnamefont {I.}~\bibnamefont {Staude}}, \bibinfo
  {author} {\bibfnamefont {A.~E.}\ \bibnamefont {Miroshnichenko}}, \bibinfo
  {author} {\bibfnamefont {D.~N.}\ \bibnamefont {Neshev}}, \bibinfo {author}
  {\bibfnamefont {A.~A.}\ \bibnamefont {Fedyanin}},\ and\ \bibinfo {author}
  {\bibfnamefont {Y.~S.}\ \bibnamefont {Kivshar}},\ }\bibfield  {title}
  {\bibinfo {title} {Ultrafast all-optical switching with magnetic resonances
  in nonlinear dielectric nanostructures},\ }\href@noop {} {\bibfield
  {journal} {\bibinfo  {journal} {Nano letters}\ }\textbf {\bibinfo {volume}
  {15}},\ \bibinfo {pages} {6985} (\bibinfo {year} {2015})}\BibitemShut
  {NoStop}%
\bibitem [{\citenamefont {Vernon}\ \emph {et~al.}(2017)\citenamefont {Vernon},
  \citenamefont {Menotti}, \citenamefont {Tison}, \citenamefont {Steidle},
  \citenamefont {Fanto}, \citenamefont {Thomas}, \citenamefont {Preble},
  \citenamefont {Smith}, \citenamefont {Alsing}, \citenamefont {Liscidini},\
  and\ \citenamefont {Sipe}}]{theorymzirings}%
  \BibitemOpen
  \bibfield  {author} {\bibinfo {author} {\bibfnamefont {Z.}~\bibnamefont
  {Vernon}}, \bibinfo {author} {\bibfnamefont {M.}~\bibnamefont {Menotti}},
  \bibinfo {author} {\bibfnamefont {C.~C.}\ \bibnamefont {Tison}}, \bibinfo
  {author} {\bibfnamefont {J.~A.}\ \bibnamefont {Steidle}}, \bibinfo {author}
  {\bibfnamefont {M.~L.}\ \bibnamefont {Fanto}}, \bibinfo {author}
  {\bibfnamefont {P.~M.}\ \bibnamefont {Thomas}}, \bibinfo {author}
  {\bibfnamefont {S.~F.}\ \bibnamefont {Preble}}, \bibinfo {author}
  {\bibfnamefont {A.~M.}\ \bibnamefont {Smith}}, \bibinfo {author}
  {\bibfnamefont {P.~M.}\ \bibnamefont {Alsing}}, \bibinfo {author}
  {\bibfnamefont {M.}~\bibnamefont {Liscidini}},\ and\ \bibinfo {author}
  {\bibfnamefont {J.~E.}\ \bibnamefont {Sipe}},\ }\bibfield  {title} {\bibinfo
  {title} {{Truly unentangled photon pairs without spectral filtering}},\
  }\href {https://doi.org/10.1364/OL.42.003638} {\bibfield  {journal} {\bibinfo
   {journal} {Opt. Lett.}\ }\textbf {\bibinfo {volume} {42}},\ \bibinfo {pages}
  {3638} (\bibinfo {year} {2017})},\ \Eprint {https://arxiv.org/abs/1703.10626}
  {arXiv:1703.10626} \BibitemShut {NoStop}%
\bibitem [{\citenamefont {Christensen}\ \emph {et~al.}(2018)\citenamefont
  {Christensen}, \citenamefont {Koefoed}, \citenamefont {Rottwitt},\ and\
  \citenamefont {McKinstrie}}]{theorydualpulse}%
  \BibitemOpen
  \bibfield  {author} {\bibinfo {author} {\bibfnamefont {J.~B.}\ \bibnamefont
  {Christensen}}, \bibinfo {author} {\bibfnamefont {J.~G.}\ \bibnamefont
  {Koefoed}}, \bibinfo {author} {\bibfnamefont {K.}~\bibnamefont {Rottwitt}},\
  and\ \bibinfo {author} {\bibfnamefont {C.~J.}\ \bibnamefont {McKinstrie}},\
  }\bibfield  {title} {\bibinfo {title} {{Engineering spectrally unentangled
  photon pairs from nonlinear microring resonators by pump manipulation}},\
  }\href {https://doi.org/10.1364/OL.43.000859} {\bibfield  {journal} {\bibinfo
   {journal} {Opt. Lett.}\ }\textbf {\bibinfo {volume} {43}},\ \bibinfo {pages}
  {859} (\bibinfo {year} {2018})}\BibitemShut {NoStop}%
\bibitem [{\citenamefont {Liu}\ \emph {et~al.}(2020)\citenamefont {Liu},
  \citenamefont {Wu}, \citenamefont {Gu}, \citenamefont {Kong}, \citenamefont
  {Yu}, \citenamefont {Ge}, \citenamefont {Cai}, \citenamefont {Qiang},
  \citenamefont {Wu}, \citenamefont {Yang},\ and\ \citenamefont
  {Others}}]{mzirings}%
  \BibitemOpen
  \bibfield  {author} {\bibinfo {author} {\bibfnamefont {Y.}~\bibnamefont
  {Liu}}, \bibinfo {author} {\bibfnamefont {C.}~\bibnamefont {Wu}}, \bibinfo
  {author} {\bibfnamefont {X.}~\bibnamefont {Gu}}, \bibinfo {author}
  {\bibfnamefont {Y.}~\bibnamefont {Kong}}, \bibinfo {author} {\bibfnamefont
  {X.}~\bibnamefont {Yu}}, \bibinfo {author} {\bibfnamefont {R.}~\bibnamefont
  {Ge}}, \bibinfo {author} {\bibfnamefont {X.}~\bibnamefont {Cai}}, \bibinfo
  {author} {\bibfnamefont {X.}~\bibnamefont {Qiang}}, \bibinfo {author}
  {\bibfnamefont {J.}~\bibnamefont {Wu}}, \bibinfo {author} {\bibfnamefont
  {X.}~\bibnamefont {Yang}},\ and\ \bibinfo {author} {\bibnamefont {Others}},\
  }\bibfield  {title} {\bibinfo {title} {{High-spectral-purity photon
  generation from a dual-interferometer-coupled silicon microring}},\
  }\href@noop {} {\bibfield  {journal} {\bibinfo  {journal} {Optics Letters}\
  }\textbf {\bibinfo {volume} {45}},\ \bibinfo {pages} {73} (\bibinfo {year}
  {2020})}\BibitemShut {NoStop}%
\bibitem [{\citenamefont {Burridge}\ \emph {et~al.}(2020)\citenamefont
  {Burridge}, \citenamefont {Faruque}, \citenamefont {Rarity},\ and\
  \citenamefont {Barreto}}]{dualpulse}%
  \BibitemOpen
  \bibfield  {author} {\bibinfo {author} {\bibfnamefont {B.~M.}\ \bibnamefont
  {Burridge}}, \bibinfo {author} {\bibfnamefont {I.~I.}\ \bibnamefont
  {Faruque}}, \bibinfo {author} {\bibfnamefont {J.~G.}\ \bibnamefont
  {Rarity}},\ and\ \bibinfo {author} {\bibfnamefont {J.}~\bibnamefont
  {Barreto}},\ }\bibfield  {title} {\bibinfo {title} {{High spectro-temporal
  purity single-photons from silicon micro-racetrack resonators using a
  dual-pulse configuration}},\ }\href@noop {} {\bibfield  {journal} {\bibinfo
  {journal} {Optics Letters}\ }\textbf {\bibinfo {volume} {45}},\ \bibinfo
  {pages} {4048} (\bibinfo {year} {2020})}\BibitemShut {NoStop}%
\bibitem [{\citenamefont {Paesani}\ \emph {et~al.}(2020)\citenamefont
  {Paesani}, \citenamefont {Borghi}, \citenamefont {Signorini}, \citenamefont
  {Ma\"inos}, \citenamefont {Pavesi},\ and\ \citenamefont
  {Laing}}]{intermodalpurity}%
  \BibitemOpen
  \bibfield  {author} {\bibinfo {author} {\bibfnamefont {S.}~\bibnamefont
  {Paesani}}, \bibinfo {author} {\bibfnamefont {M.}~\bibnamefont {Borghi}},
  \bibinfo {author} {\bibfnamefont {S.}~\bibnamefont {Signorini}}, \bibinfo
  {author} {\bibfnamefont {A.}~\bibnamefont {Ma\"inos}}, \bibinfo {author}
  {\bibfnamefont {L.}~\bibnamefont {Pavesi}},\ and\ \bibinfo {author}
  {\bibfnamefont {A.}~\bibnamefont {Laing}},\ }\bibfield  {title} {\bibinfo
  {title} {{Near-ideal spontaneous photon sources in silicon quantum
  photonics}},\ }\href@noop {} {\bibfield  {journal} {\bibinfo  {journal}
  {Nature communications}\ }\textbf {\bibinfo {volume} {11}},\ \bibinfo {pages}
  {1} (\bibinfo {year} {2020})}\BibitemShut {NoStop}%
\bibitem [{\citenamefont {Silverstone}\ \emph {et~al.}(2014)\citenamefont
  {Silverstone}, \citenamefont {Bonneau}, \citenamefont {Ohira}, \citenamefont
  {Suzuki}, \citenamefont {Yoshida}, \citenamefont {Iizuka}, \citenamefont
  {Ezaki}, \citenamefont {Natarajan}, \citenamefont {Tanner}, \citenamefont
  {Hadfield},\ and\ \citenamefont {Others}}]{Silverstonespiralnoise}%
  \BibitemOpen
  \bibfield  {author} {\bibinfo {author} {\bibfnamefont {J.~W.}\ \bibnamefont
  {Silverstone}}, \bibinfo {author} {\bibfnamefont {D.}~\bibnamefont
  {Bonneau}}, \bibinfo {author} {\bibfnamefont {K.}~\bibnamefont {Ohira}},
  \bibinfo {author} {\bibfnamefont {N.}~\bibnamefont {Suzuki}}, \bibinfo
  {author} {\bibfnamefont {H.}~\bibnamefont {Yoshida}}, \bibinfo {author}
  {\bibfnamefont {N.}~\bibnamefont {Iizuka}}, \bibinfo {author} {\bibfnamefont
  {M.}~\bibnamefont {Ezaki}}, \bibinfo {author} {\bibfnamefont {C.~M.}\
  \bibnamefont {Natarajan}}, \bibinfo {author} {\bibfnamefont {M.~G.}\
  \bibnamefont {Tanner}}, \bibinfo {author} {\bibfnamefont {R.~H.}\
  \bibnamefont {Hadfield}},\ and\ \bibinfo {author} {\bibnamefont {Others}},\
  }\bibfield  {title} {\bibinfo {title} {{On-chip quantum interference between
  silicon photon-pair sources}},\ }\href@noop {} {\bibfield  {journal}
  {\bibinfo  {journal} {Nature Photonics}\ }\textbf {\bibinfo {volume} {8}},\
  \bibinfo {pages} {104} (\bibinfo {year} {2014})}\BibitemShut {NoStop}%
\bibitem [{\citenamefont {Paesani}\ \emph {et~al.}(2019)\citenamefont
  {Paesani}, \citenamefont {Ding}, \citenamefont {Santagati}, \citenamefont
  {Chakhmakhchyan}, \citenamefont {Vigliar}, \citenamefont {Rottwitt},
  \citenamefont {Oxenl{\o}we}, \citenamefont {Wang}, \citenamefont {Thompson},\
  and\ \citenamefont {Laing}}]{samplingstates}%
  \BibitemOpen
  \bibfield  {author} {\bibinfo {author} {\bibfnamefont {S.}~\bibnamefont
  {Paesani}}, \bibinfo {author} {\bibfnamefont {Y.}~\bibnamefont {Ding}},
  \bibinfo {author} {\bibfnamefont {R.}~\bibnamefont {Santagati}}, \bibinfo
  {author} {\bibfnamefont {L.}~\bibnamefont {Chakhmakhchyan}}, \bibinfo
  {author} {\bibfnamefont {C.}~\bibnamefont {Vigliar}}, \bibinfo {author}
  {\bibfnamefont {K.}~\bibnamefont {Rottwitt}}, \bibinfo {author}
  {\bibfnamefont {L.~K.}\ \bibnamefont {Oxenl{\o}we}}, \bibinfo {author}
  {\bibfnamefont {J.}~\bibnamefont {Wang}}, \bibinfo {author} {\bibfnamefont
  {M.~G.}\ \bibnamefont {Thompson}},\ and\ \bibinfo {author} {\bibfnamefont
  {A.}~\bibnamefont {Laing}},\ }\bibfield  {title} {\bibinfo {title}
  {{Generation and sampling of quantum states of light in a silicon chip}},\
  }\href@noop {} {\bibfield  {journal} {\bibinfo  {journal} {Nature Physics}\
  }\textbf {\bibinfo {volume} {15}},\ \bibinfo {pages} {925} (\bibinfo {year}
  {2019})}\BibitemShut {NoStop}%
\bibitem [{\citenamefont {Llewellyn}\ \emph {et~al.}(2020)\citenamefont
  {Llewellyn}, \citenamefont {Ding}, \citenamefont {Faruque}, \citenamefont
  {Paesani}, \citenamefont {Bacco}, \citenamefont {Santagati}, \citenamefont
  {Qian}, \citenamefont {Li}, \citenamefont {Xiao}, \citenamefont {Huber},\
  and\ \citenamefont {Others}}]{teleport}%
  \BibitemOpen
  \bibfield  {author} {\bibinfo {author} {\bibfnamefont {D.}~\bibnamefont
  {Llewellyn}}, \bibinfo {author} {\bibfnamefont {Y.}~\bibnamefont {Ding}},
  \bibinfo {author} {\bibfnamefont {I.~I.}\ \bibnamefont {Faruque}}, \bibinfo
  {author} {\bibfnamefont {S.}~\bibnamefont {Paesani}}, \bibinfo {author}
  {\bibfnamefont {D.}~\bibnamefont {Bacco}}, \bibinfo {author} {\bibfnamefont
  {R.}~\bibnamefont {Santagati}}, \bibinfo {author} {\bibfnamefont {Y.-J.}\
  \bibnamefont {Qian}}, \bibinfo {author} {\bibfnamefont {Y.}~\bibnamefont
  {Li}}, \bibinfo {author} {\bibfnamefont {Y.-F.}\ \bibnamefont {Xiao}},
  \bibinfo {author} {\bibfnamefont {M.}~\bibnamefont {Huber}},\ and\ \bibinfo
  {author} {\bibnamefont {Others}},\ }\bibfield  {title} {\bibinfo {title}
  {{Chip-to-chip quantum teleportation and multi-photon entanglement in
  silicon}},\ }\href@noop {} {\bibfield  {journal} {\bibinfo  {journal} {Nature
  Physics}\ }\textbf {\bibinfo {volume} {16}},\ \bibinfo {pages} {148}
  (\bibinfo {year} {2020})}\BibitemShut {NoStop}%
\bibitem [{\citenamefont {Faruque}\ \emph {et~al.}(2018)\citenamefont
  {Faruque}, \citenamefont {Sinclair}, \citenamefont {Bonneau}, \citenamefont
  {Rarity},\ and\ \citenamefont {Thompson}}]{imadrings}%
  \BibitemOpen
  \bibfield  {author} {\bibinfo {author} {\bibfnamefont {I.~I.}\ \bibnamefont
  {Faruque}}, \bibinfo {author} {\bibfnamefont {G.~F.}\ \bibnamefont
  {Sinclair}}, \bibinfo {author} {\bibfnamefont {D.}~\bibnamefont {Bonneau}},
  \bibinfo {author} {\bibfnamefont {J.~G.}\ \bibnamefont {Rarity}},\ and\
  \bibinfo {author} {\bibfnamefont {M.~G.}\ \bibnamefont {Thompson}},\
  }\bibfield  {title} {\bibinfo {title} {{On-chip quantum interference with
  heralded photons from two independent micro-ring resonator sources in silicon
  photonics}},\ }\href {https://doi.org/10.1364/OE.26.020379} {\bibfield
  {journal} {\bibinfo  {journal} {Opt. Express}\ }\textbf {\bibinfo {volume}
  {26}},\ \bibinfo {pages} {20379} (\bibinfo {year} {2018})}\BibitemShut
  {NoStop}%
\bibitem [{\citenamefont {Silverstone}\ \emph {et~al.}(2015)\citenamefont
  {Silverstone}, \citenamefont {Santagati}, \citenamefont {Bonneau},
  \citenamefont {Strain}, \citenamefont {Sorel}, \citenamefont {O'Brien},\ and\
  \citenamefont {Thompson}}]{silverstonerings}%
  \BibitemOpen
  \bibfield  {author} {\bibinfo {author} {\bibfnamefont {J.~W.}\ \bibnamefont
  {Silverstone}}, \bibinfo {author} {\bibfnamefont {R.}~\bibnamefont
  {Santagati}}, \bibinfo {author} {\bibfnamefont {D.}~\bibnamefont {Bonneau}},
  \bibinfo {author} {\bibfnamefont {M.~J.}\ \bibnamefont {Strain}}, \bibinfo
  {author} {\bibfnamefont {M.}~\bibnamefont {Sorel}}, \bibinfo {author}
  {\bibfnamefont {J.~L.}\ \bibnamefont {O'Brien}},\ and\ \bibinfo {author}
  {\bibfnamefont {M.~G.}\ \bibnamefont {Thompson}},\ }\bibfield  {title}
  {\bibinfo {title} {{Qubit entanglement between ring-resonator photon-pair
  sources on a silicon chip}},\ }\href@noop {} {\bibfield  {journal} {\bibinfo
  {journal} {Nature communications}\ }\textbf {\bibinfo {volume} {6}},\
  \bibinfo {pages} {1} (\bibinfo {year} {2015})}\BibitemShut {NoStop}%
\bibitem [{\citenamefont {Wang}\ \emph {et~al.}(2018)\citenamefont {Wang},
  \citenamefont {Paesani}, \citenamefont {Ding}, \citenamefont {Santagati},
  \citenamefont {Skrzypczyk}, \citenamefont {Salavrakos}, \citenamefont {Tura},
  \citenamefont {Augusiak}, \citenamefont {Man{\v{c}}inska}, \citenamefont
  {Bacco},\ and\ \citenamefont {Others}}]{highdimensional}%
  \BibitemOpen
  \bibfield  {author} {\bibinfo {author} {\bibfnamefont {J.}~\bibnamefont
  {Wang}}, \bibinfo {author} {\bibfnamefont {S.}~\bibnamefont {Paesani}},
  \bibinfo {author} {\bibfnamefont {Y.}~\bibnamefont {Ding}}, \bibinfo {author}
  {\bibfnamefont {R.}~\bibnamefont {Santagati}}, \bibinfo {author}
  {\bibfnamefont {P.}~\bibnamefont {Skrzypczyk}}, \bibinfo {author}
  {\bibfnamefont {A.}~\bibnamefont {Salavrakos}}, \bibinfo {author}
  {\bibfnamefont {J.}~\bibnamefont {Tura}}, \bibinfo {author} {\bibfnamefont
  {R.}~\bibnamefont {Augusiak}}, \bibinfo {author} {\bibfnamefont
  {L.}~\bibnamefont {Man{\v{c}}inska}}, \bibinfo {author} {\bibfnamefont
  {D.}~\bibnamefont {Bacco}},\ and\ \bibinfo {author} {\bibnamefont {Others}},\
  }\bibfield  {title} {\bibinfo {title} {{Multidimensional quantum entanglement
  with large-scale integrated optics}},\ }\href@noop {} {\bibfield  {journal}
  {\bibinfo  {journal} {Science}\ }\textbf {\bibinfo {volume} {360}},\ \bibinfo
  {pages} {285} (\bibinfo {year} {2018})}\BibitemShut {NoStop}%
\bibitem [{\citenamefont {Qiang}\ \emph {et~al.}(2018)\citenamefont {Qiang},
  \citenamefont {Zhou}, \citenamefont {Wang}, \citenamefont {Wilkes},
  \citenamefont {Loke}, \citenamefont {O'Gara}, \citenamefont {Kling},
  \citenamefont {Marshall}, \citenamefont {Santagati}, \citenamefont {Ralph},\
  and\ \citenamefont {Others}}]{arbtwoqubit}%
  \BibitemOpen
  \bibfield  {author} {\bibinfo {author} {\bibfnamefont {X.}~\bibnamefont
  {Qiang}}, \bibinfo {author} {\bibfnamefont {X.}~\bibnamefont {Zhou}},
  \bibinfo {author} {\bibfnamefont {J.}~\bibnamefont {Wang}}, \bibinfo {author}
  {\bibfnamefont {C.~M.}\ \bibnamefont {Wilkes}}, \bibinfo {author}
  {\bibfnamefont {T.}~\bibnamefont {Loke}}, \bibinfo {author} {\bibfnamefont
  {S.}~\bibnamefont {O'Gara}}, \bibinfo {author} {\bibfnamefont
  {L.}~\bibnamefont {Kling}}, \bibinfo {author} {\bibfnamefont {G.~D.}\
  \bibnamefont {Marshall}}, \bibinfo {author} {\bibfnamefont {R.}~\bibnamefont
  {Santagati}}, \bibinfo {author} {\bibfnamefont {T.~C.}\ \bibnamefont
  {Ralph}},\ and\ \bibinfo {author} {\bibnamefont {Others}},\ }\bibfield
  {title} {\bibinfo {title} {{Large-scale silicon quantum photonics
  implementing arbitrary two-qubit processing}},\ }\href@noop {} {\bibfield
  {journal} {\bibinfo  {journal} {Nature photonics}\ }\textbf {\bibinfo
  {volume} {12}},\ \bibinfo {pages} {534} (\bibinfo {year} {2018})}\BibitemShut
  {NoStop}%
\bibitem [{\citenamefont {Santagati}\ \emph {et~al.}(2017)\citenamefont
  {Santagati}, \citenamefont {Silverstone}, \citenamefont {Strain},
  \citenamefont {Sorel}, \citenamefont {Miki}, \citenamefont {Yamashita},
  \citenamefont {Fujiwara}, \citenamefont {Sasaki}, \citenamefont {Terai},
  \citenamefont {Tanner},\ and\ \citenamefont {Others}}]{twoqubitentangle}%
  \BibitemOpen
  \bibfield  {author} {\bibinfo {author} {\bibfnamefont {R.}~\bibnamefont
  {Santagati}}, \bibinfo {author} {\bibfnamefont {J.~W.}\ \bibnamefont
  {Silverstone}}, \bibinfo {author} {\bibfnamefont {M.~J.}\ \bibnamefont
  {Strain}}, \bibinfo {author} {\bibfnamefont {M.}~\bibnamefont {Sorel}},
  \bibinfo {author} {\bibfnamefont {S.}~\bibnamefont {Miki}}, \bibinfo {author}
  {\bibfnamefont {T.}~\bibnamefont {Yamashita}}, \bibinfo {author}
  {\bibfnamefont {M.}~\bibnamefont {Fujiwara}}, \bibinfo {author}
  {\bibfnamefont {M.}~\bibnamefont {Sasaki}}, \bibinfo {author} {\bibfnamefont
  {H.}~\bibnamefont {Terai}}, \bibinfo {author} {\bibfnamefont {M.~G.}\
  \bibnamefont {Tanner}},\ and\ \bibinfo {author} {\bibnamefont {Others}},\
  }\bibfield  {title} {\bibinfo {title} {{Silicon photonic processor of
  two-qubit entangling quantum logic}},\ }\href@noop {} {\bibfield  {journal}
  {\bibinfo  {journal} {Journal of Optics}\ }\textbf {\bibinfo {volume} {19}},\
  \bibinfo {pages} {114006} (\bibinfo {year} {2017})}\BibitemShut {NoStop}%
\bibitem [{\citenamefont {Harris}\ \emph {et~al.}(2014)\citenamefont {Harris},
  \citenamefont {Grassani}, \citenamefont {Simbula}, \citenamefont {Pant},
  \citenamefont {Galli}, \citenamefont {Baehr-Jones}, \citenamefont {Hochberg},
  \citenamefont {Englund}, \citenamefont {Bajoni},\ and\ \citenamefont
  {Galland}}]{filteringintegrated}%
  \BibitemOpen
  \bibfield  {author} {\bibinfo {author} {\bibfnamefont {N.~C.}\ \bibnamefont
  {Harris}}, \bibinfo {author} {\bibfnamefont {D.}~\bibnamefont {Grassani}},
  \bibinfo {author} {\bibfnamefont {A.}~\bibnamefont {Simbula}}, \bibinfo
  {author} {\bibfnamefont {M.}~\bibnamefont {Pant}}, \bibinfo {author}
  {\bibfnamefont {M.}~\bibnamefont {Galli}}, \bibinfo {author} {\bibfnamefont
  {T.}~\bibnamefont {Baehr-Jones}}, \bibinfo {author} {\bibfnamefont
  {M.}~\bibnamefont {Hochberg}}, \bibinfo {author} {\bibfnamefont
  {D.}~\bibnamefont {Englund}}, \bibinfo {author} {\bibfnamefont
  {D.}~\bibnamefont {Bajoni}},\ and\ \bibinfo {author} {\bibfnamefont
  {C.}~\bibnamefont {Galland}},\ }\bibfield  {title} {\bibinfo {title}
  {{Integrated source of spectrally filtered correlated photons for large-scale
  quantum photonic systems}},\ }\href@noop {} {\bibfield  {journal} {\bibinfo
  {journal} {Physical Review X}\ }\textbf {\bibinfo {volume} {4}},\ \bibinfo
  {pages} {41047} (\bibinfo {year} {2014})}\BibitemShut {NoStop}%
\bibitem [{\citenamefont {Wang}\ \emph {et~al.}(2020)\citenamefont {Wang},
  \citenamefont {Sciarrino}, \citenamefont {Laing},\ and\ \citenamefont
  {Thompson}}]{integratedquantumphotonic}%
  \BibitemOpen
  \bibfield  {author} {\bibinfo {author} {\bibfnamefont {J.}~\bibnamefont
  {Wang}}, \bibinfo {author} {\bibfnamefont {F.}~\bibnamefont {Sciarrino}},
  \bibinfo {author} {\bibfnamefont {A.}~\bibnamefont {Laing}},\ and\ \bibinfo
  {author} {\bibfnamefont {M.~G.}\ \bibnamefont {Thompson}},\ }\bibfield
  {title} {\bibinfo {title} {{Integrated photonic quantum technologies}},\
  }\href@noop {} {\bibfield  {journal} {\bibinfo  {journal} {Nature Photonics}\
  }\textbf {\bibinfo {volume} {14}},\ \bibinfo {pages} {273} (\bibinfo {year}
  {2020})}\BibitemShut {NoStop}%
\bibitem [{\citenamefont {Seok}\ \emph {et~al.}(2019)\citenamefont {Seok},
  \citenamefont {Kwon}, \citenamefont {Henriksson}, \citenamefont {Luo},\ and\
  \citenamefont {Wu}}]{largescaleswitches}%
  \BibitemOpen
  \bibfield  {author} {\bibinfo {author} {\bibfnamefont {T.~J.}\ \bibnamefont
  {Seok}}, \bibinfo {author} {\bibfnamefont {K.}~\bibnamefont {Kwon}}, \bibinfo
  {author} {\bibfnamefont {J.}~\bibnamefont {Henriksson}}, \bibinfo {author}
  {\bibfnamefont {J.}~\bibnamefont {Luo}},\ and\ \bibinfo {author}
  {\bibfnamefont {M.~C.}\ \bibnamefont {Wu}},\ }\bibfield  {title} {\bibinfo
  {title} {240$\times$ 240 wafer-scale silicon photonic switches},\ }in\
  \href@noop {} {\emph {\bibinfo {booktitle} {Optical Fiber Communication
  Conference}}}\ (\bibinfo {organization} {Optical Society of America},\
  \bibinfo {year} {2019})\ pp.\ \bibinfo {pages} {Th1E----5}\BibitemShut
  {NoStop}%
\bibitem [{\citenamefont {P{\'{e}}rez}\ \emph {et~al.}(2017)\citenamefont
  {P{\'{e}}rez}, \citenamefont {Gasulla}, \citenamefont {Crudgington},
  \citenamefont {Thomson}, \citenamefont {Khokhar}, \citenamefont {Li},
  \citenamefont {Cao}, \citenamefont {Mashanovich},\ and\ \citenamefont
  {Capmany}}]{photonicprocessor}%
  \BibitemOpen
  \bibfield  {author} {\bibinfo {author} {\bibfnamefont {D.}~\bibnamefont
  {P{\'{e}}rez}}, \bibinfo {author} {\bibfnamefont {I.}~\bibnamefont
  {Gasulla}}, \bibinfo {author} {\bibfnamefont {L.}~\bibnamefont
  {Crudgington}}, \bibinfo {author} {\bibfnamefont {D.~J.}\ \bibnamefont
  {Thomson}}, \bibinfo {author} {\bibfnamefont {A.~Z.}\ \bibnamefont
  {Khokhar}}, \bibinfo {author} {\bibfnamefont {K.}~\bibnamefont {Li}},
  \bibinfo {author} {\bibfnamefont {W.}~\bibnamefont {Cao}}, \bibinfo {author}
  {\bibfnamefont {G.~Z.}\ \bibnamefont {Mashanovich}},\ and\ \bibinfo {author}
  {\bibfnamefont {J.}~\bibnamefont {Capmany}},\ }\bibfield  {title} {\bibinfo
  {title} {{Multipurpose silicon photonics signal processor core}},\
  }\href@noop {} {\bibfield  {journal} {\bibinfo  {journal} {Nature
  communications}\ }\textbf {\bibinfo {volume} {8}},\ \bibinfo {pages} {1}
  (\bibinfo {year} {2017})}\BibitemShut {NoStop}%
\bibitem [{\citenamefont {Moody}\ \emph {et~al.}(2020)\citenamefont {Moody},
  \citenamefont {Chang}, \citenamefont {Steiner},\ and\ \citenamefont
  {Bowers}}]{nonlineargeneration}%
  \BibitemOpen
  \bibfield  {author} {\bibinfo {author} {\bibfnamefont {G.}~\bibnamefont
  {Moody}}, \bibinfo {author} {\bibfnamefont {L.}~\bibnamefont {Chang}},
  \bibinfo {author} {\bibfnamefont {T.~J.}\ \bibnamefont {Steiner}},\ and\
  \bibinfo {author} {\bibfnamefont {J.~E.}\ \bibnamefont {Bowers}},\ }\bibfield
   {title} {\bibinfo {title} {{Chip-scale nonlinear photonics for quantum light
  generation}},\ }\href@noop {} {\bibfield  {journal} {\bibinfo  {journal} {AVS
  Quantum Science}\ }\textbf {\bibinfo {volume} {2}},\ \bibinfo {pages} {41702}
  (\bibinfo {year} {2020})}\BibitemShut {NoStop}%
\bibitem [{\citenamefont {Signorini}\ and\ \citenamefont
  {Pavesi}(2020)}]{singlephotonsources}%
  \BibitemOpen
  \bibfield  {author} {\bibinfo {author} {\bibfnamefont {S.}~\bibnamefont
  {Signorini}}\ and\ \bibinfo {author} {\bibfnamefont {L.}~\bibnamefont
  {Pavesi}},\ }\bibfield  {title} {\bibinfo {title} {{On-chip heralded single
  photon sources}},\ }\href@noop {} {\bibfield  {journal} {\bibinfo  {journal}
  {AVS Quantum Science}\ }\textbf {\bibinfo {volume} {2}},\ \bibinfo {pages}
  {41701} (\bibinfo {year} {2020})}\BibitemShut {NoStop}%
\bibitem [{asy(2012)}]{asyfields}%
  \BibitemOpen
  \bibfield  {title} {\bibinfo {title} {{Asymptotic fields for a Hamiltonian
  treatment of nonlinear electromagnetic phenomena}},\ }\bibfield  {journal}
  {\bibinfo  {journal} {Physical Review A - Atomic, Molecular, and Optical
  Physics}\ }\textbf {\bibinfo {volume} {85}},\ \href
  {https://doi.org/10.1103/PhysRevA.85.013833} {10.1103/PhysRevA.85.013833}
  (\bibinfo {year} {2012})\BibitemShut {NoStop}%
\bibitem [{\citenamefont {Boyd}(2008)}]{nonlinoptics}%
  \BibitemOpen
  \bibfield  {author} {\bibinfo {author} {\bibfnamefont {R.~W.}\ \bibnamefont
  {Boyd}},\ }\href@noop {} {\emph {\bibinfo {title} {{Nonlinear Optics}}}},\
  \bibinfo {edition} {3rd}\ ed.\ (\bibinfo  {publisher} {Academic Press,
  Inc.},\ \bibinfo {year} {2008})\ pp.\ \bibinfo {pages} {77--78}\BibitemShut
  {NoStop}%
\bibitem [{\citenamefont {Husko}\ \emph {et~al.}(2013)\citenamefont {Husko},
  \citenamefont {Clark}, \citenamefont {Collins}, \citenamefont {{De Rossi}},
  \citenamefont {Combri{\'{e}}}, \citenamefont {Lehoucq}, \citenamefont {Rey},
  \citenamefont {Krauss}, \citenamefont {Xiong},\ and\ \citenamefont
  {Eggleton}}]{photonabsorption}%
  \BibitemOpen
  \bibfield  {author} {\bibinfo {author} {\bibfnamefont {C.~A.}\ \bibnamefont
  {Husko}}, \bibinfo {author} {\bibfnamefont {A.~S.}\ \bibnamefont {Clark}},
  \bibinfo {author} {\bibfnamefont {M.~J.}\ \bibnamefont {Collins}}, \bibinfo
  {author} {\bibfnamefont {A.}~\bibnamefont {{De Rossi}}}, \bibinfo {author}
  {\bibfnamefont {S.}~\bibnamefont {Combri{\'{e}}}}, \bibinfo {author}
  {\bibfnamefont {G.}~\bibnamefont {Lehoucq}}, \bibinfo {author} {\bibfnamefont
  {I.~H.}\ \bibnamefont {Rey}}, \bibinfo {author} {\bibfnamefont {T.~F.}\
  \bibnamefont {Krauss}}, \bibinfo {author} {\bibfnamefont {C.}~\bibnamefont
  {Xiong}},\ and\ \bibinfo {author} {\bibfnamefont {B.~J.}\ \bibnamefont
  {Eggleton}},\ }\bibfield  {title} {\bibinfo {title} {{Multi-photon absorption
  limits to heralded single photon sources}},\ }\href@noop {} {\bibfield
  {journal} {\bibinfo  {journal} {Scientific reports}\ }\textbf {\bibinfo
  {volume} {3}},\ \bibinfo {pages} {1} (\bibinfo {year} {2013})}\BibitemShut
  {NoStop}%
\bibitem [{\citenamefont {Yang}\ \emph {et~al.}(2008)\citenamefont {Yang},
  \citenamefont {Liscidini},\ and\ \citenamefont {Sipe}}]{photongentheory}%
  \BibitemOpen
  \bibfield  {author} {\bibinfo {author} {\bibfnamefont {Z.}~\bibnamefont
  {Yang}}, \bibinfo {author} {\bibfnamefont {M.}~\bibnamefont {Liscidini}},\
  and\ \bibinfo {author} {\bibfnamefont {J.~E.}\ \bibnamefont {Sipe}},\
  }\bibfield  {title} {\bibinfo {title} {Spontaneous parametric down-conversion
  in waveguides: a backward heisenberg picture approach},\ }\href@noop {}
  {\bibfield  {journal} {\bibinfo  {journal} {Physical Review A}\ }\textbf
  {\bibinfo {volume} {77}},\ \bibinfo {pages} {033808} (\bibinfo {year}
  {2008})}\BibitemShut {NoStop}%
\bibitem [{\citenamefont {Faruque}\ \emph {et~al.}(2021)\citenamefont
  {Faruque}, \citenamefont {Burridge}, \citenamefont {Borghi}, \citenamefont
  {Barreto},\ and\ \citenamefont {Rarity}}]{SpET}%
  \BibitemOpen
  \bibfield  {author} {\bibinfo {author} {\bibfnamefont {I.~I.}\ \bibnamefont
  {Faruque}}, \bibinfo {author} {\bibfnamefont {B.}~\bibnamefont {Burridge}},
  \bibinfo {author} {\bibfnamefont {M.}~\bibnamefont {Borghi}}, \bibinfo
  {author} {\bibfnamefont {J.}~\bibnamefont {Barreto}},\ and\ \bibinfo {author}
  {\bibfnamefont {J.}~\bibnamefont {Rarity}},\ }\bibfield  {title} {\bibinfo
  {title} {Phase tomography of spontaneously emitted photon-pairs},\ }in\
  \href@noop {} {\emph {\bibinfo {booktitle} {CLEO: QELS\_Fundamental
  Science}}}\ (\bibinfo {organization} {Optical Society of America},\ \bibinfo
  {year} {2021})\ pp.\ \bibinfo {pages} {FF2I--1}\BibitemShut {NoStop}%
\bibitem [{\citenamefont {Wehner}\ \emph {et~al.}(2018)\citenamefont {Wehner},
  \citenamefont {Elkouss},\ and\ \citenamefont {Hanson}}]{qinternet}%
  \BibitemOpen
  \bibfield  {author} {\bibinfo {author} {\bibfnamefont {S.}~\bibnamefont
  {Wehner}}, \bibinfo {author} {\bibfnamefont {D.}~\bibnamefont {Elkouss}},\
  and\ \bibinfo {author} {\bibfnamefont {R.}~\bibnamefont {Hanson}},\
  }\bibfield  {title} {\bibinfo {title} {Quantum internet: A vision for the
  road ahead},\ }\href@noop {} {\bibfield  {journal} {\bibinfo  {journal}
  {Science}\ }\textbf {\bibinfo {volume} {362}},\ \bibinfo {pages} {eaam9288}
  (\bibinfo {year} {2018})}\BibitemShut {NoStop}%
\bibitem [{\citenamefont {Bartolucci}\ \emph {et~al.}(2021)\citenamefont
  {Bartolucci}, \citenamefont {Birchall}, \citenamefont {Bombin}, \citenamefont
  {Cable}, \citenamefont {Dawson}, \citenamefont {Gimeno-Segovia},
  \citenamefont {Johnston}, \citenamefont {Kieling}, \citenamefont {Nickerson},
  \citenamefont {Pant} \emph {et~al.}}]{qcompfusion}%
  \BibitemOpen
  \bibfield  {author} {\bibinfo {author} {\bibfnamefont {S.}~\bibnamefont
  {Bartolucci}}, \bibinfo {author} {\bibfnamefont {P.}~\bibnamefont
  {Birchall}}, \bibinfo {author} {\bibfnamefont {H.}~\bibnamefont {Bombin}},
  \bibinfo {author} {\bibfnamefont {H.}~\bibnamefont {Cable}}, \bibinfo
  {author} {\bibfnamefont {C.}~\bibnamefont {Dawson}}, \bibinfo {author}
  {\bibfnamefont {M.}~\bibnamefont {Gimeno-Segovia}}, \bibinfo {author}
  {\bibfnamefont {E.}~\bibnamefont {Johnston}}, \bibinfo {author}
  {\bibfnamefont {K.}~\bibnamefont {Kieling}}, \bibinfo {author} {\bibfnamefont
  {N.}~\bibnamefont {Nickerson}}, \bibinfo {author} {\bibfnamefont
  {M.}~\bibnamefont {Pant}}, \emph {et~al.},\ }\bibfield  {title} {\bibinfo
  {title} {Fusion-based quantum computation},\ }\href@noop {} {\bibfield
  {journal} {\bibinfo  {journal} {arXiv preprint arXiv:2101.09310}\ } (\bibinfo
  {year} {2021})}\BibitemShut {NoStop}%
\bibitem [{\citenamefont {Faruque}\ \emph {et~al.}(2019)\citenamefont
  {Faruque}, \citenamefont {Sinclair}, \citenamefont {Bonneau}, \citenamefont
  {Ono}, \citenamefont {Silberhorn}, \citenamefont {Thompson},\ and\
  \citenamefont {Rarity}}]{imadg2}%
  \BibitemOpen
  \bibfield  {author} {\bibinfo {author} {\bibfnamefont {I.~I.}\ \bibnamefont
  {Faruque}}, \bibinfo {author} {\bibfnamefont {G.~F.}\ \bibnamefont
  {Sinclair}}, \bibinfo {author} {\bibfnamefont {D.}~\bibnamefont {Bonneau}},
  \bibinfo {author} {\bibfnamefont {T.}~\bibnamefont {Ono}}, \bibinfo {author}
  {\bibfnamefont {C.}~\bibnamefont {Silberhorn}}, \bibinfo {author}
  {\bibfnamefont {M.~G.}\ \bibnamefont {Thompson}},\ and\ \bibinfo {author}
  {\bibfnamefont {J.~G.}\ \bibnamefont {Rarity}},\ }\bibfield  {title}
  {\bibinfo {title} {Estimating the indistinguishability of heralded single
  photons using second-order correlation},\ }\href@noop {} {\bibfield
  {journal} {\bibinfo  {journal} {Physical Review Applied}\ }\textbf {\bibinfo
  {volume} {12}},\ \bibinfo {pages} {054029} (\bibinfo {year}
  {2019})}\BibitemShut {NoStop}%
\bibitem [{\citenamefont {Wang}\ \emph {et~al.}(2022)\citenamefont {Wang},
  \citenamefont {Fargas~Cabanillas}, \citenamefont {Kramnik}, \citenamefont
  {Ramesh}, \citenamefont {Buchbinder}, \citenamefont {Kumar}, \citenamefont
  {Stojanovic},\ and\ \citenamefont {Popovic}}]{filterpresource}%
  \BibitemOpen
  \bibfield  {author} {\bibinfo {author} {\bibfnamefont {I.}~\bibnamefont
  {Wang}}, \bibinfo {author} {\bibfnamefont {J.}~\bibnamefont
  {Fargas~Cabanillas}}, \bibinfo {author} {\bibfnamefont {D.}~\bibnamefont
  {Kramnik}}, \bibinfo {author} {\bibfnamefont {A.}~\bibnamefont {Ramesh}},
  \bibinfo {author} {\bibfnamefont {S.}~\bibnamefont {Buchbinder}}, \bibinfo
  {author} {\bibfnamefont {P.}~\bibnamefont {Kumar}}, \bibinfo {author}
  {\bibfnamefont {V.}~\bibnamefont {Stojanovic}},\ and\ \bibinfo {author}
  {\bibfnamefont {M.}~\bibnamefont {Popovic}},\ }\bibfield  {title} {\bibinfo
  {title} {Toward quantum electronic-photonic systems-on-chip: a monolithic
  source of quantum-correlated photons with integrated frequency locking
  electronics and pump rejection},\ }in\ \href@noop {} {\emph {\bibinfo
  {booktitle} {CLEO: Science and Innovations}}}\ (\bibinfo {organization}
  {Optical Society of America},\ \bibinfo {year} {2022})\ pp.\ \bibinfo {pages}
  {SM3N--2}\BibitemShut {NoStop}%
\bibitem [{\citenamefont {Liscidini}\ and\ \citenamefont {Sipe}(2013)}]{SET}%
  \BibitemOpen
  \bibfield  {author} {\bibinfo {author} {\bibfnamefont {M.}~\bibnamefont
  {Liscidini}}\ and\ \bibinfo {author} {\bibfnamefont {J.}~\bibnamefont
  {Sipe}},\ }\bibfield  {title} {\bibinfo {title} {Stimulated emission
  tomography},\ }\href@noop {} {\bibfield  {journal} {\bibinfo  {journal}
  {Physical review letters}\ }\textbf {\bibinfo {volume} {111}},\ \bibinfo
  {pages} {193602} (\bibinfo {year} {2013})}\BibitemShut {NoStop}%
\bibitem [{\citenamefont {Davis}\ \emph {et~al.}(2017)\citenamefont {Davis},
  \citenamefont {Saulnier}, \citenamefont {Karpi{\'n}ski},\ and\ \citenamefont
  {Smith}}]{photonspectrum}%
  \BibitemOpen
  \bibfield  {author} {\bibinfo {author} {\bibfnamefont {A.~O.}\ \bibnamefont
  {Davis}}, \bibinfo {author} {\bibfnamefont {P.~M.}\ \bibnamefont {Saulnier}},
  \bibinfo {author} {\bibfnamefont {M.}~\bibnamefont {Karpi{\'n}ski}},\ and\
  \bibinfo {author} {\bibfnamefont {B.~J.}\ \bibnamefont {Smith}},\ }\bibfield
  {title} {\bibinfo {title} {Pulsed single-photon spectrometer by
  frequency-to-time mapping using chirped fiber bragg gratings},\ }\href@noop
  {} {\bibfield  {journal} {\bibinfo  {journal} {Optics Express}\ }\textbf
  {\bibinfo {volume} {25}},\ \bibinfo {pages} {12804} (\bibinfo {year}
  {2017})}\BibitemShut {NoStop}%
\bibitem [{\citenamefont {Eckstein}\ \emph {et~al.}(2011)\citenamefont
  {Eckstein}, \citenamefont {Christ}, \citenamefont {Mosley},\ and\
  \citenamefont {Silberhorn}}]{realg2}%
  \BibitemOpen
  \bibfield  {author} {\bibinfo {author} {\bibfnamefont {A.}~\bibnamefont
  {Eckstein}}, \bibinfo {author} {\bibfnamefont {A.}~\bibnamefont {Christ}},
  \bibinfo {author} {\bibfnamefont {P.~J.}\ \bibnamefont {Mosley}},\ and\
  \bibinfo {author} {\bibfnamefont {C.}~\bibnamefont {Silberhorn}},\ }\bibfield
   {title} {\bibinfo {title} {Realistic g (2) measurement of a pdc source with
  single photon detectors in the presence of background},\ }\href@noop {}
  {\bibfield  {journal} {\bibinfo  {journal} {physica status solidi c}\
  }\textbf {\bibinfo {volume} {8}},\ \bibinfo {pages} {1216} (\bibinfo {year}
  {2011})}\BibitemShut {NoStop}%
\bibitem [{\citenamefont {Lee}\ \emph {et~al.}(2019)\citenamefont {Lee},
  \citenamefont {Lee}, \citenamefont {Kim},\ and\ \citenamefont
  {Ju}}]{noisesources}%
  \BibitemOpen
  \bibfield  {author} {\bibinfo {author} {\bibfnamefont {J.-M.}\ \bibnamefont
  {Lee}}, \bibinfo {author} {\bibfnamefont {W.-J.}\ \bibnamefont {Lee}},
  \bibinfo {author} {\bibfnamefont {M.-S.}\ \bibnamefont {Kim}},\ and\ \bibinfo
  {author} {\bibfnamefont {J.~J.}\ \bibnamefont {Ju}},\ }\bibfield  {title}
  {\bibinfo {title} {Noise filtering for highly correlated photon pairs from
  silicon waveguides},\ }\href@noop {} {\bibfield  {journal} {\bibinfo
  {journal} {Journal of Lightwave Technology}\ }\textbf {\bibinfo {volume}
  {37}},\ \bibinfo {pages} {5428} (\bibinfo {year} {2019})}\BibitemShut
  {NoStop}%
\bibitem [{\citenamefont {Clemmen}\ \emph {et~al.}(2012)\citenamefont
  {Clemmen}, \citenamefont {Perret}, \citenamefont {Safioui}, \citenamefont
  {Bogaerts}, \citenamefont {Baets}, \citenamefont {Gorza}, \citenamefont
  {Emplit},\ and\ \citenamefont {Massar}}]{inelasticscattering}%
  \BibitemOpen
  \bibfield  {author} {\bibinfo {author} {\bibfnamefont {S.}~\bibnamefont
  {Clemmen}}, \bibinfo {author} {\bibfnamefont {A.}~\bibnamefont {Perret}},
  \bibinfo {author} {\bibfnamefont {J.}~\bibnamefont {Safioui}}, \bibinfo
  {author} {\bibfnamefont {W.}~\bibnamefont {Bogaerts}}, \bibinfo {author}
  {\bibfnamefont {R.}~\bibnamefont {Baets}}, \bibinfo {author} {\bibfnamefont
  {S.-P.}\ \bibnamefont {Gorza}}, \bibinfo {author} {\bibfnamefont
  {P.}~\bibnamefont {Emplit}},\ and\ \bibinfo {author} {\bibfnamefont
  {S.}~\bibnamefont {Massar}},\ }\bibfield  {title} {\bibinfo {title}
  {Low-power inelastic light scattering at small detunings in silicon wire
  waveguides at telecom wavelengths},\ }\href@noop {} {\bibfield  {journal}
  {\bibinfo  {journal} {JOSA B}\ }\textbf {\bibinfo {volume} {29}},\ \bibinfo
  {pages} {1977} (\bibinfo {year} {2012})}\BibitemShut {NoStop}%
\bibitem [{\citenamefont {Menotti}\ \emph {et~al.}(2019)\citenamefont
  {Menotti}, \citenamefont {Morrison}, \citenamefont {Tan}, \citenamefont
  {Vernon}, \citenamefont {Sipe},\ and\ \citenamefont
  {Liscidini}}]{linearuncouple}%
  \BibitemOpen
  \bibfield  {author} {\bibinfo {author} {\bibfnamefont {M.}~\bibnamefont
  {Menotti}}, \bibinfo {author} {\bibfnamefont {B.}~\bibnamefont {Morrison}},
  \bibinfo {author} {\bibfnamefont {K.}~\bibnamefont {Tan}}, \bibinfo {author}
  {\bibfnamefont {Z.}~\bibnamefont {Vernon}}, \bibinfo {author} {\bibfnamefont
  {J.}~\bibnamefont {Sipe}},\ and\ \bibinfo {author} {\bibfnamefont
  {M.}~\bibnamefont {Liscidini}},\ }\bibfield  {title} {\bibinfo {title}
  {Nonlinear coupling of linearly uncoupled resonators},\ }\href@noop {}
  {\bibfield  {journal} {\bibinfo  {journal} {Physical review letters}\
  }\textbf {\bibinfo {volume} {122}},\ \bibinfo {pages} {013904} (\bibinfo
  {year} {2019})}\BibitemShut {NoStop}%
\bibitem [{\citenamefont {Sabattoli}\ \emph {et~al.}(2021)\citenamefont
  {Sabattoli}, \citenamefont {El~Dirani}, \citenamefont {Youssef},
  \citenamefont {Garrisi}, \citenamefont {Grassani}, \citenamefont {Zatti},
  \citenamefont {Petit-Etienne}, \citenamefont {Pargon}, \citenamefont {Sipe},
  \citenamefont {Liscidini} \emph {et~al.}}]{linearuncouple_parasitic}%
  \BibitemOpen
  \bibfield  {author} {\bibinfo {author} {\bibfnamefont {F.~A.}\ \bibnamefont
  {Sabattoli}}, \bibinfo {author} {\bibfnamefont {H.}~\bibnamefont
  {El~Dirani}}, \bibinfo {author} {\bibfnamefont {L.}~\bibnamefont {Youssef}},
  \bibinfo {author} {\bibfnamefont {F.}~\bibnamefont {Garrisi}}, \bibinfo
  {author} {\bibfnamefont {D.}~\bibnamefont {Grassani}}, \bibinfo {author}
  {\bibfnamefont {L.}~\bibnamefont {Zatti}}, \bibinfo {author} {\bibfnamefont
  {C.}~\bibnamefont {Petit-Etienne}}, \bibinfo {author} {\bibfnamefont
  {E.}~\bibnamefont {Pargon}}, \bibinfo {author} {\bibfnamefont
  {J.}~\bibnamefont {Sipe}}, \bibinfo {author} {\bibfnamefont {M.}~\bibnamefont
  {Liscidini}}, \emph {et~al.},\ }\bibfield  {title} {\bibinfo {title}
  {Suppression of parasitic nonlinear processes in spontaneous four-wave mixing
  with linearly uncoupled resonators},\ }\href@noop {} {\bibfield  {journal}
  {\bibinfo  {journal} {Physical Review Letters}\ }\textbf {\bibinfo {volume}
  {127}},\ \bibinfo {pages} {033901} (\bibinfo {year} {2021})}\BibitemShut
  {NoStop}%
\end{thebibliography}%

\end{document}